%% file: weak-mem-survey-masterfile.tex
\documentclass[acmsmall,nonacm]{acmart}

\usepackage{silence}
\usepackage{wrapfig}

\usepackage{xspace}
\usepackage{transparent}
\usepackage[inline]{enumitem}
\usepackage{survey-definitions}
\usepackage{stmaryrd}
\usepackage{mathtools}

\usepackage{opsem-survey}

\usepackage{anyfontsize}

\usepackage[most]{tcolorbox}

\setcitestyle{nosort}

\newcommand{\ie}{i.e.,\xspace}
\newcommand{\eg}{e.g.,\xspace}
\newcommand{\etal}{et al.\xspace}

\newcounter{mmdefncounter}

\newcommand{\exmpheading}[1]
{
  \refstepcounter{examplecounter}%
  \textbf{Example \arabic{section}.\arabic{examplecounter} (#1).}%
}

\newcounter{examplecounter}
\numberwithin{examplecounter}{section}

\newenvironment{ourbox}
{
  \begin{tcolorbox}[arc=0mm, boxrule=0mm, breakable]
}
{
  \end{tcolorbox}
}

\newenvironment{exmp}[1][]
{
  \begin{tcolorbox}[float, floatplacement=ht, arc=0mm, boxrule=0mm]
  \exmpheading{#1}
}
{
  \end{tcolorbox}
}

\newenvironment{multiexmp}
{
  \begin{tcolorbox}[float, floatplacement=t, arc=0mm, boxrule=0mm]
}
{
  \end{tcolorbox}
}

\newcommand{\exmpsep}{\vskip 3mm}

\newcommand{\ourparagraph}[1]{\noindent\textbf{#1.}}
\renewcommand{\ourparagraph}[1]{\paragraph{\textbf{#1}}}

\usepackage{doi,mathpartir,natbib,tabularx,ulem,xcolor}
\setcitestyle{square,numbers,citesep={,}} 
\usetikzlibrary{quotes,arrows.meta}

\title{Weak Memory Model Formalisms: Introduction and Survey}

\thanks{This is the pre-peer-reviewed version of the article: 
``Weak memory model formalisms: Introduction and survey'', \emph{Concurrency and Computation: Practice and Experience}, 38(2), 2026,
which has been published in final form at \texttt{doi.org/10.1002/cpe.70484}. This article may be used for non-commercial purposes in accordance with Wiley Terms and Conditions for Use of Self-Archived Versions.}

\authorsaddresses{Authors' addresses: Su (roger.c.su@proton.me), Colvin (r.colvin@uq.edu.au). August 2025.}

\author{Roger C. Su}
	\affiliation{
		\institution{School of Computing, Australian National University}
		\city{Canberra} 
		\country{Australia}
	}
	\email{roger.su@anu.edu.au}
	\orcid{0000-0003-0176-3264}

\author{Robert J. Colvin}
	\affiliation{
		\institution{Defence Science and Technology Group}
		\city{Brisbane}
		\country{Australia}
	}
	\affiliation{
		\institution{School of Electrical Engineering and Computer Science, University of Queensland}
		\city{Brisbane}
		\country{Australia}
	}
	\email{r.colvin@uq.edu.au}






\begin{document}

\begin{abstract}




Memory consistency models define the order in which accesses to shared memory in a concurrent system may be observed to occur. Such models are a necessity since \emph{program order} is not a reliable indicator of \emph{execution order}, due to microarchitectural features or compiler transformations. Concurrent programming, already a challenging task, is thus made even harder when weak memory effects must be addressed.  A rigorous specification of weak memory models is therefore essential to make this problem tractable for developers of safety- and security-critical, low-level software.

In this paper we survey the field of formalisations of weak memory models, including their specification, their effects on execution, and tools and inference systems for reasoning about code.  To assist the discussion we also provide an introduction to two styles of formal representation found commonly in the literature (using a much simplified version of Intel's x86 as the example): a step-by-step construction of traces of the system (\emph{operational semantics}); and with respect to relations between memory events (\emph{axiomatic semantics}).  The survey covers some long-standing hardware features that lead to observable weak behaviours, a description of historical developments in practice and in theory, an overview of computability and complexity results, and outlines current and future directions in the field.

\end{abstract}

\maketitle

\section{Introduction}\labelsect{intro}
\labelsect{introduction}

\OMIT{
Programmers of low-level concurrent systems (\eg operating systems or library data structures) need to address the behaviour of the shared memory on the target architecture, so that the processes can communicate correctly via shared-memory accesses.
This is a challenging task, as architectures (single- or multi-core) do not guarantee to execute instructions in the order specified, and nor do optimising compilers guarantee to maintain the order of the program statements in their output code.
This adds to the difficulties to concurrent programming, which is already complicated even without the potential out-of-order factor above.

To address this extra complication, processor manufacturers and language designers define
a \emph{memory model}, which is a minimal set of constraints about the behaviour of interacting concurrent processes, typically with respect to the order of shared-memory accesses.
}

Programmers of low-level concurrent systems code -- \eg operating systems or library data structures -- need to address the behaviour of the
target architecture; specifically, how communication occurs via accesses to shared memory.  This is nontrivial, as architectures (single- or multi-core) do not
guarantee to execute instructions in the order specified, and nor do compilers guarantee to maintain the order of the program statements in the generated
code.  This can add to the difficulties already associated with concurrency under the standard, sequential interpretation.

For example, consider an Arm or RISC-V processor executing assembly corresponding to the following higher-level code,
where one process 
writes data $\xd$ to a shared variable $\xx$ and then signals this change 
to a concurrent process (or processes) via a shared ``$\xflag$'' variable.
The reading process spins waiting for the flag to be set and then reads the data from $\xx$ into its local state (variable $\xreg$).
\begin{equation}
\labeleqn{first-example}
\mathtt{
	x \asgn d \scomp flag \asgn 1
	\quad
	\pl
	\quad
	(\While flag \neq 1 \Do \{\}) \scomp r \asgn x
}
\end{equation}
The processor executing the first process
can potentially execute the two instructions out of order, since, from the perspective of a single processor, the order is irrelevant
to the final values of the two variables.  
As such, the flag may be set \emph{before} $\xd$ is written to $\xx$, and thus the second process may receive an unintended value.
In fact, by the same justification, even if the first process performs its instructions in the order specified,
the second processor may execute the load of $\xx$ before the loop is resolved (a type of \emph{speculative execution}), again
resulting in an erroneous value for $\xx$.

To address this extra complication introduced by architectures and compiler optimisations, processor vendors and language designers define
a ``memory model'', a minimal set of constraints about the behaviour of interacting concurrent processes, typically with respect to the
order of accesses of shared memory.  
If those constraints are too weak to enforce the logic of an algorithm, programmers may
inject ``fences'' or related mechanisms that can restore the standard, sequential behaviour of a (segment of a) program. 
For instance, the use of 
\emph{release/acquire} constraints is a well-known programming pattern for addressing weak memory model effects \cite{ReleaseConsistency90}.
A release fence, `$\mathtt{rel}$',
(in \Clang, \texttt{atomic\_thread\_fence(memory\_order\_release)})
can be injected into the first process of \refeqn{first-example},
$
\mathtt{
	x \asgn 1 \scomp \mathtt{rel} \scomp flag \asgn 1
}
$.
In tandem with an acquire fence after the loop of the second process these
ensure the intended value for $\mathtt{x}$ will be read.
However, excessive use of such mechanisms comes with performance penalties,
which can be a significant issue for low-level systems code~\cite{ritson-16}.  

\OMIT{
restore order by injecting ``fences'' or related mechanisms that can restore the standard, sequential behaviour of a program. 
However, excessive use of such mechanisms may hinder the performance,
which can be a significant issue for low-level programming \cite{ritson-16}.  
}

Since the early days of multiprocessors (a system consisting of multiple processors) to the present age of multicore processors (each processor consisting of multiple cores), the literature on weak memory models has developed
in academia, the commercial sector, and standardisation committees.
The issue affects hardware designers, compiler writers, application programmers, and language designers~\cite{adve-ghara-96-ieee-comput,adve-boehm-10-commun-acm}.
As yet, no consensus on a particular \emph{formal} representation has been settled upon,
leaving programmers -- in particular those requiring formal assurance for the code -- with many choices.
The company Arm has led the way in formalising their architecture and robustly investigating their design, including with publicly accessible tool support.
Their formal model is validated on hardware through the use of thousands of small programs
called ``litmus tests''.
However, the mechanisms for inferring properties of programs using their formalisation are complex, even for simple programs;
as a result, many other representations have been considered.

In this paper, we survey these diverse \emph{formalisms} and \emph{tools} that have been proposed for representing weak memory models.
A formal specification of a memory model should preferably match the existing intuitions about processor behaviours and compiler optimisations.
Furthermore, it should support an easy enumeration of the consequences: a guide that feeds into the software developers' effort. 
Though this paper is primarily aimed at researchers with a background in formal methods,
we consider the field from the perspective of a practitioner:
\begin{itemize}
\item
\emph{How to specify a memory model?}
(hardware designers; language designers)%
;
\item
\emph{How to understand the effect of a memory model on a program?}  
(compiler writers; application programmers)%
; and
\item
\emph{How to formally verify that a program is correct under a particular memory model?} 
(application programmers -- specifically for secure or safety-critical systems)%
.
\end{itemize}
Each of these questions has many potential answers, depending on the level of microarchitectural detail vs. abstraction desired,
level of tool support, and program properties under question.

\paragraph{Scope and Limitations.}
In this survey we attempt to cover the majority of common formalisations of modern hardware and software weak memory models, with reference to historical developments where there are clear impacts on 
current practice.  We are particularly focussed on formalisations that lead to tools and techniques for inferring properties of programs,
\ie from a developers perspective.

We aim to trace the historical development of processor design from the 1960s, through theoretical
developments in the 1990s and up to the present day, 
which includes regular contributions in major publishing venues and
refinements of commercial specifications. 
As a result, we were unable to find a methodical sequence of search terms of the type used
in a recent survey on programming language memory models \cite{mois-21}. 
Hence, we do not claim full coverage, but intend that the range of works surveyed is indicative of the state of the field.

We do not specifically consider issues related to hardware or compiler construction:
we assume that microarchitectural implementations of instruction set architectures faithfully match their specifications, and that compiler transformations correctly preserve their intended software memory model in the underlying architecture.
Of course, 
this is not always the case, with problems being discovered in hardware~\cite{alglave-14,ReadAfterReadHazardNotice2011} 
and potential compiler mappings \cite{TriCheck}.

%


\paragraph{Outline.} 
We first introduce terminology and conventions (\refsect{prelim-terminology}), and informally introduce \emph{litmus tests} (\refsect{prelim-litmus}), which are small 
programs designed to concisely show weak memory effects.
We then briefly review aspects of processors that give rise to weak memory effects (\refsect{history}), and give an overview of prominent weak memory models, past and present (\refsect{prelim}).

As an introduction to formalisations and the range of representations that have been considered we 
present an
\emph{operational} (\refsect{operat}) representation based on 
\emph{write buffers} (as defined in \cite{x86-TSO}),
and
and an \emph{axiomatic} (\refsect{declar}) view
that abstracts from architectural details and instead focusses on properties of traces of behaviours
(based on work best exemplified in \cite{alglave-14}).
We use a representative fragment of Intel's x86 as a running example of a memory model,
and use the \emph{store buffering} litmus test (\refexmp{store-buffer}) as the running example program.
Because examples are traditionally the best way to explain weak memory effects, we keep the 
them separate from the main text, in highlighted boxes, to assist readers looking for instances over details.

In \refsect{formalisms} we survey and compare operational, axiomatic, and other representations from the literature.
Verification methods building on formalisations are surveyed in \refsect{tools-logic},
and tool support (and associated algorithmic problems) is surveyed in \refsect{tools}.
We suggest ways forward for unifying research into and properties of weak memory models in \refsect{future-properties},
and outline how formalisations intersect with other challenges in software verification in \refsect{future-adapt}.

\subsection{Terminology and Conventions}\labelsect{prelim-terminology}

We clarify some of the terminology and naming conventions used in this paper, as there is significant overloading of terms in the literature.
Our survey covers both low-level and high-level, abstract concepts, but we tend to choose the more abstract terminology.

\emph{Processes and Variables.}
A shared-memory concurrent system consists of a set of \emph{processes} (covering \emph{threads} and \emph{cores}) and a \emph{shared memory}.
The shared memory consists of \emph{shared variables} (covering \emph{locations}).
Each process can access \emph{local variables} (covering \emph{registers}), which are not accessible by other processes.

\emph{Instructions and Events.}
Processes execute \emph{instructions} such as \Loads and \Stores.
Instructions may generate \Reads and \Writes,
which are \emph{events} (or \emph{operations}) on the shared memory. %
Some instructions generate more than one event;
these instructions are called \emph{atomic primitives} or \emph{read-modify-write}s (RMWs), and examples include the
``test-and-set'' (TAS) and ``compare-and-swap'' (CAS) instructions.

By convention we let $v$ be a value, typically an integer or byte representation.
We write $x \asgn v$ to denote the instruction that stores the value $v$ to the shared variable $x$.
In the literature, such a \Store may be written as ``$x.\Store(v)$,''
or in an assembler language such as x86 as 
``\texttt{MOV [rx] v}'', where \texttt{rx} is a register that holds the location in shared memory associated with a program variable $x$.
As above we tend to use the more abstract syntax, which may be straightforwardly converted to other styles.
(Similarly, $r \asgn x$ denotes a \emph{load} of shared variable $x$ into register $r$, sometimes written $r = x.\Load()$
or ``\texttt{MOV r [rx]}''.)

\emph{\SeqCst.}
The standard, intuitive meaning of a concurrent program is a set of \emph{interleavings}. 
Each interleaving is a sequence of all operations, such that if operation $\aca$ is specified in the program to run before operation $\acb$, then $\aca$ will be executed before $\acb$.
When a concurrent program is viewed this way, we say that it is executed under \emph{\SeqCst}.
As we will see in \refsect{history}, this is not generally guaranteed on real processors or by compilers.

\emph{Memory Models.}
The word ``model'' is heavily overloaded and potentially ambiguous.%
\footnote{
For example, a ``memory model'' can mean the structure of the memory address space. This usage, unrelated to concurrency, occurs in microarchitectures~\citep[p.72]{gonzalez-11} as well as in \Clang programming~\citep{lepigre-22,memarian-16}. In the setting of \Clang, a memory model in this sense is sometimes called a ``memory object model.''}
In this paper, a \emph{memory model}\footnote{A \emph{memory model} is also called \emph{memory consistency}, \emph{consistency model} and \emph{memory consistency model}~\citep{book-consistency-2}.}
denotes the underlying set of possible behaviours of every concurrent program. 
\SeqCst is a memory model, and in fact the most simple one.
When a program exhibits behaviour that is not compatible with \SeqCst, then the program is said to run under a \emph{weak} (or \emph{relaxed}) memory model.

\OMIT{
\emph{Formalisms.}
Traditionally, semantics of programs are often classified as operational, denotational, or axiomatic~\cite{Jones-18}. 
In the literature on weak memory models, formalisms are often classified as either ``axiomatic'' or 
``operational'', which we indicate as appropriate.
We additionally distinguish formalisms on whether 
the underlying \emph{representation} used to understand behaviours due to weak memory
is ``mechanistic'' -- that is, explains behaviours using an abstract structure that corresponds to some degree 
to a mechanism used by real processors.
}


\subsection{Litmus Tests}\labelsect{prelim-litmus}
\labelsect{litmus-tests}

\newcommand{\tick}{$\green{\checkmark}$}
\newcommand{\cross}{$\red{\times}$}
\newcommand{\SCmm}{\textsc{sc}}

\newcommand{\refexmps}[2]{\refexmp{#1} and~\refexmp{#2}}
\newcommand{\refexmpsr}[2]{Examples~\ref{exmp:#1}--\ref{exmp:#2}}

Weak memory models are traditionally understood using small programs that highlight particular unexpected behaviours,
typically called \emph{litmus tests}.
Given some initial state of the shared and local variables,
one can ``probe'' a hardware or software memory model by checking whether a final state is \emph{reachable} for a given litmus test.
\refexmpsr{store-buffer}{mp} give several
litmus tests, including the \emph{store buffering} litmus test, which we use as a running example,
and
is one of the most common litmus tests referred to in the literature~\cite{TutorialARMandPOWER,mador-haim-11}.
\reftable{litmuses-vs-models} summarises the relationships between the key litmus tests and the common memory models.

\newcommand{\refexmppage}[1]{p.\pageref{exmp:#1}}

\newcommand{\refexmpintable}[1]{(\refex{#1}, \refexmppage{#1})}

\begin{table}[h]
  \centering
  \caption{Comparison of the major memory models on common litmus tests}
  \labeltable{litmuses-vs-models}
  \begin{tabular}{|cc||c|c|c|c|c|}\hline
    &  &   SC   &  x86   &  Arm   & RISC-V & C     \\\hline\hline
	  SB & \refexmpintable{store-buffer} & \cross & \tick  & \tick  & \tick  & \tick \\\hline
	  LB  & \refexmpintable{load-buffer}& \cross & \cross & \tick  & \tick  & \tick \\\hline
	  MP & \refexmpintable{mp} & \cross & \cross & \tick  & \tick  & \tick \\\hline
	  IRIW & \refexmpintable{iriw}& \cross & \cross & \cross & \cross & \tick \\\hline
	  LB+ctrls & \refexmpintable{oota}& \cross & \cross & \cross & \cross & \tick \\\hline
  \end{tabular}
\end{table}

\begin{multiexmp}
By convention, here and throughout $x,y,z$ denote shared variables, $r, r_1, r_2$ denote local variables,
and all variables are initially 0.
The final states of interest are enclosed in braces $\{ \ldots \}$ after the pieces of code.
We call the left-hand side process Process A, and the right-hand side process Process B. 

\exmpsep

\exmpheading{Store Buffering (SB)}
\labelexmp{store-buffer}

\begin{wrapfigure}[6]{R}{0.36\textwidth}
	\vskip -4mm
	\begin{tabular}{c}
				\begin{tabular}{l||l}
					$x := 1$ ; ~&~  $y := 1$ ; \\
					$r_1 := y$ ~&~ $r_2 := x$
				\end{tabular}
				\\
				$\{r_1 = 0 \land r_2 = 0\}$ \\
			\begin{tabular}{ll}
				SC & \cross \\ x86, Arm, RISC-V, C & \tick
			\end{tabular}
			\end{tabular}
\end{wrapfigure}

In the \emph{store buffering} litmus test Process A stores 1 to $x$ and then loads the value of $y$. 
Process B does the converse by storing 1 to $y$ and then loading the value of $x$.
Under \SeqCst, the last instruction to take place must be either of the two \Loads. The \Load that takes place last must return the value 1, so $r_1$ and $r_2$ can never both be 0.
However, on most modern processors, it is possible to reach a final state where both $r_1$ and $r_2$ are 0.
On the software-level, \Clang compilers may also choose to reorder the instructions, making this final state reachable
regardless of the hardware.

\exmpsep

\exmpheading{Load Buffering}
\labelexmp{load-buffer}

\begin{wrapfigure}[6]{r}{0.31\textwidth}
	\vspace{-6mm}
	\begin{tabular}{c}
		\begin{tabular}{l||l}
			$r_1 := y$ ; ~&~ $r_2 := x$ ;  \\
			$x := 1$ ~&~ $y := 1$
		\end{tabular}
		\\
		$\{r_1 = 1 \land r_2 = 1\}$ \\
		\begin{tabular}{ll}
			SC, x86 & \cross \\ Arm, RISC-V, C & \tick
		\end{tabular}
	\end{tabular}
\end{wrapfigure}

Here the order of instructions in SB are swapped,
and similarly under \SeqCst it is not possible for both \Loads to return the value 1.
This outcome is disallowed on x86 architectures,
but is however reachable on Arm or RISC-V processors,
and, as above, a compiler may choose to reorder the instructions regardless.

\exmpsep

\exmpheading{Message Passing}
\labelexmp{mp}

\begin{wrapfigure}[6]{r}{0.31\textwidth}
	\vspace{-6mm}
	\begin{tabular}{c}
		\begin{tabular}{l||l}
			$x := 1$ ; ~&~ $r_1 := y$ ; \\
			$y := 1$ ~&~ $r_2 := x$ \\
		\end{tabular}
		\\
		$\{r_1 = 1 \land r_2 = 0\}$ \\
		\begin{tabular}{ll}
			SC, x86 & \cross \\ Arm, RISC-V, C & \tick
		\end{tabular}
	\end{tabular}
\end{wrapfigure}

In this common pattern Process A stores 1 to $x$, and then ``informs'' Process B by storing 1 to $y$, signalling that an update to $x$ has occurred
(similarly to \refeqn{first-example}).
Under \SeqCst or on x86 processors, if Process B observes $y = 1$, then it may infer that $x = 1$ also.
However, this need not be the case on Arm, RISC-V, or in \Clang.

\end{multiexmp}

Several tools exist for automatically generating litmus tests from a weak memory model specification \cite{mador-haim-2010,lustig-2017-inproc,Alglave-Litmus,AlglaveBlog}.

\OMIT{
In many cases, the discrepancies between the ``expected'' behaviour and the observed behaviour can
be understood in terms of the processor pipelines or other specific microarchitectural features
(see \refsect{history} and \refexmp{iriw}).
Another well-known, complex phenomenon is the \emph{out of thin air} behaviour,
where under some circumstances circularity in dependencies can lead to values being assigned to variables
that were never explicitly written in the text (see \refsect{prelim-soft} and \refexmp{oota}).
Such behaviours do not manifest on hardware, but can theoretically arise from compiler optimisations, and hence
must be addressed in software memory models.
}

\input{history}

\input{mechanistic}

\input{declarative}

\input{formalisations}

\input{other-semantics}

\input{verification-and-tools}

\input{future-directions}

\input{conclusions}

\begin{acks}
The authors thank Scott Heiner, Graeme Smith and the anonymous reviewers for their detailed and insightful feedback on the paper.
This project was funded by Department of Defence and administered through the Advanced Strategic Capabilities Accelerator.
\end{acks}

\hbadness = 10000

\bibliographystyle{acm-ref-tmp}
\bibliography{refs-survey}

\end{document}

%% file: history.tex

\section{The Development of Processors}
\labelsect{history}

\newcommand{\val}[1]{\mathsf{#1}}
\newcommand{\vala}{\val{a}}
\newcommand{\valb}{\val{b}}
\newcommand{\valc}{\val{c}}
\newcommand{\vald}{\val{d}}
\newcommand{\instrRX}{\mathtt{r_1 \asgn x}}
\newcommand{\instrRY}{\mathtt{r_2 \asgn y}}
\newcommand{\instrRR}{\mathtt{r_3 \asgn r_1 + r_2}}

\newcommand{\plain}[1]{$#1$}
\newcommand{\issued}[1]{\uline{$#1$}}
\newcommand{\completed}[1]{\sout{$#1$}}
\newcommand{\dependent}[1]{\mathbf{#1}}

To contextualise how weak memory behaviours arise we sketch some historical developments in processor design. 
We begin with one aspect of processor microarchitectures which may lead to weak memory behaviours: instruction-level parallelism within the pipeline (\refsect{pipelining}). 
We then discuss
fence instructions (and other mechanisms) that can be used to restore order if necessary (\refsect{fences}).
We also outline other microarchitectural features such as forwarding (\refsect{forwarding}) and (the lack of) multicopy atomicity (\refsect{multicopy-atomicity}). 
Milestones in processor design relevant to the development of weak memory models are outlined in \reftable{processors-chronology}.

\begin{table}[t]
  \caption{Milestones in processor design relevant to the development of weak memory models}
  \labeltable{processors-chronology}
  \vspace{-4mm}
  \begin{tabularx}{\textwidth}{|lX|}
    \hline
    \textbf{1963}&
The CDC 6600 \cite{CDC6600} significantly improved processor speed by using a ``scoreboard'' to track which registers and locations are being accessed by instructions in the pipeline; instructions that have no dependencies may be parallelised.
    \\\hline
    \textbf{1967}&
The first implementation of Tomasulo's algorithm~\citep{tomasulo-67} was Model 91 in IBM's System/360 family~\citep{IBM-360-91}, which improved the utilisation of multiple arithmetic units.
    \\\hline
    \textbf{1980s}&
    The Stanford DASH multiprocessor~\citep{DASH} brought together much existing research and influenced later processors.
    \\\hline
    \textbf{1990s}&
    Speculative execution was introduced by Intel's Pentium family of processors (a type of x86 architecture). 
    \\\hline
    \textbf{2001}&
    IBM's POWER4 was the first commercially available multicore processor. It includes a cache system with even weaker guarantees than those given by pipeline reordering, and is not multicopy-atomic (\refsect{lmca}).
    \\\hline
    \textbf{2011}&
    RISC-V was released as an openly available specification \cite{RISC-V}, with a memory model based on Release Consistency (see \refsect{prelim-early}).
    \\\hline
    \textbf{2017}&
    Based on research outcomes, as described in \cite{alglave-14}, the Arm specification explicitly became multicopy-atomic \cite{pulte-17}.
    \\\hline
  \end{tabularx}
\end{table}

\subsection{Instruction-Level Parallelism}
\labelsect{pipelining}
\labelsect{pipelines}

\OMIT{
To execute a piece of code,
a processor \emph{fetches} instructions from the \emph{code base}\footnote{The code base contains assembly instructions that may be pre-compiled or compiled ``just in time'' from a high-level language.}
and places them into a \emph{pipeline}, from where the instructions are decoded and executed.
}

The execution of a single instruction may involve multiple micro-steps, including to access registers, calculate values, or read and write main memory.
These latter micro-steps, which we call ``memory operations'', can include complex interactions with the cache system, or across PCI devices.  Any one of these or many other aspects of computer hardware 
\cite{hennessy2011computer,patterson2016computer,mckenney2010memory}
could potentially
influence the actual order of execution of the original sequence of compiled instructions.

Focussing at the local level, the performance of a processor can be improved 
in part by executing \emph{independent} micro-steps in parallel. 
Techniques in this vein are collectively termed \emph{instruction-level parallelism}, and have been developed since the early days of computing~\cite{rao-93}.
Straightforward cases include parallelising loads of distinct locations in memory, and parallelising independent calculations to utilise multiple arithmetic and logic units (ALUs) on a single chip.
The decision whether or not to parallelise is guided by the observable effect on the sequential semantics -- if the programmer could potentially observe the parallelism on a single chip, then it should not be allowed 
(see \refexmp{parallelize-arith}).
Early examples include the ``scoreboarding'' technique of the CDC 6600 computer \cite{CDC6600} and Tomasulo's algorithm \cite{tomasulo-67}.
Parallelism can also be employed when accessing distinct locations in memory (see \refexmp{parallelize-loads}).

\begin{multiexmp}
\exmpheading{Parallelising independent arithmetic instructions}
\labelexmp{parallelize-arith}

\begin{wrapfigure}[4]{r}{0.17\textwidth}
\vskip -7mm
\begin{minipage}{0.17\textwidth}
\[
\begin{array}{l}
r_1 \asgn \vala * \valb
\scomp
\\
r_2 \asgn \valc + \vald
\scomp
\\
r_3 \asgn r_1 + r_2
\end{array}
\]
\end{minipage}
\end{wrapfigure}

In this sequential program, where $\vala, \valb, \valc$ and $\vald$ are some integer values,
the final value of $r_3$ depends on the calculated values $r_1$ and $r_2$.
Assuming the multiplication of $\vala$ and $\valb$ is computationally expensive,
it makes sense to start calculating $\valc + \vald$ in the second instruction using an addition ALU, 
if one is available.
Meanwhile, the final instruction depends on the results of the first two, so it is effectively blocked until both
have completed. 
In whichever order the first two instructions are executed, the final state
satisfies $r_3 = \vala * \valb + \valc + \vald$, but
the overall running time is potentially reduced by executing the first two instructions in parallel.

\exmpsep

\exmpheading{Parallelising independent loads}
\labelexmp{parallelize-loads}

\begin{wrapfigure}[4]{r}{0.17\textwidth}
\vskip -7mm
\begin{minipage}{0.17\textwidth}
\[
\begin{array}{l}
r_1 \asgn x
\scomp \\
r_2 \asgn y
\scomp \\
r_3 \asgn r_1 + r_2
\end{array}
\]
\end{minipage}
\end{wrapfigure}

In this sequential program $x$ and $y$ are shared variables and $r_1,r_2,r_3$ are registers.
On a single processor
it does not matter which of $x$ or $y$ are loaded from memory first, and thus the first two instructions can be executed in parallel.
However, as in \refexmp{parallelize-arith}, the third instruction $r_3 = x+y$ can only complete after the 
values for $x$ and $y$ have been read.

\end{multiexmp}

As an example of how microarchitectural features can result in weak memory effects
we outline below
a standard view of processor pipelines.
Pipeline processing can be generally considered in three main phases.
First, instructions are \emph{fetched} from the code base into the pipeline.
Second, instructions in the pipeline are \emph{issued} to local ALUs, registers, or the main memory.
Third, when all required values have been retrieved and calculated, the instructions are \emph{completed} and removed from the pipeline.
As different instructions may take different lengths of time in the \emph{issue} and \emph{complete} phases,
they may be parallelised.

\newcommand{\instrX}{\mathtt{x \asgn 1}}
\newcommand{\instrY}{\mathtt{y \asgn 1}}
\renewcommand{\instrRY}{\mathtt{r_2 \asgn y}}
\renewcommand{\instrRX}{\mathtt{r_1 \asgn x}}

\newcommand{\fetchabbrev}{F}
\newcommand{\issueabbrev}{I}
\newcommand{\complabbrev}{C}

\begin{multiexmp}

\exmpheading{Pipelines and calculation dependencies}
\labelexmp{deps-pipeline}

\begin{wrapfigure}[11]{r}{0.5\textwidth}
\vskip -4mm
\begin{tabular}{ll}
  Phase & Pipeline
  \\\hline
  Fetch & $\instrRX$ 
  \\
  Fetch & $\instrRX$, $\instrRY$ 
  \\
  Fetch & $\instrRX$, $\instrRY$, $\instrRR$ 
  \\
  Issue & \issued{\instrRX}, $\instrRY$, $\mathtt{r_3 \asgn \dependent{r_1} + \dependent{r_2}}$ 
  \\
  Issue & \issued{\instrRX}, \issued{\instrRY}, $\mathtt{r_3 \asgn \dependent{r_1} + \dependent{r_2}}$ 
  \\
  Complete & \issued{\instrRX}, \completed{\instrRY}, $\mathtt{r_3 \asgn \dependent{r_1} + r_2}$ 
  \\
  Complete & \completed{\instrRX}, \completed{\instrRY}, $\instrRR$
  \\
  Issue & \completed{\instrRX}, \completed{\instrRY}, \issued{\instrRR}
  \\
  Complete & \completed{\instrRX}, \completed{\instrRY}, \completed{\instrRR} \\\hline
\end{tabular}
\end{wrapfigure}

The table illustrates the workings of a processor pipeline \cite{hennessy2011computer}
given 
the sequential program 
in \refexmp{parallelize-loads}
	($\instrRX \scomp \instrRY \scomp \instrRR$).
When an instruction is fetched from the code base and placed in the pipe\-line, we add this instruction to the pipe\-line column of the table.
An instruction is underlined when it is issued, and is struck out when completed.
When the value of a variable depends on incomplete instructions earlier in the pipeline, the variable is in bold font.

In this example,
the instruction $\instrRY$ completes before $\instrRX$, despite being fetched and issued later.
This could be due to a range of microarchitectural and environmental factors
-- for instance, if $y$ is already stored in the local cache (a ``cache hit'') and $x$ is not (a ``cache miss'').


\exmpsep

\exmpheading{Store buffering in the pipeline}
\labelexmp{sb-pipeline}
Recall the store buffering litmus test (\refexmp{store-buffer}), 
where Process A executes $\instrX \scomp \instrRY$,
and Process B executes $\instrY \scomp \instrRX$.
Each process represents a separate processor or core, and thus has a pipeline of its own.

In each process, the two instructions are independent.
When Process A runs in isolation, no matter in which order it carries out the two instructions, the final state always satisfies 
$x = 1 \land r_1 = 0$.
Similarly, Process B always arrives at $y = 1 \land r_2 = 0$.


However, concurrent pipelines may cause unexpected behaviours, as shown below.
(We abbreviate \emph{fetch}, \emph{issue}, and \emph{complete} by \fetchabbrev, \issueabbrev, and \complabbrev, respectively, 
and condense some steps.)

\begin{center}
\begin{tabular}{l|l||l||l|l}
	Phases & Pipeline A & Notes & Phases & Pipeline B 
	\\\hline
	\fetchabbrev, \fetchabbrev & 
	\plain{\instrX}, \plain{\instrRY} &
	&
	\fetchabbrev, \fetchabbrev & 
	\plain{\instrY}, \plain{\instrRX} \\
	\issueabbrev, \issueabbrev & 
	\issued{\instrX}, \issued{\instrRY} &
	&
	\issueabbrev, \issueabbrev & 
	\issued{\instrY}, \issued{\instrRX} \\
	\complabbrev &
	\issued{\instrX}, \completed{\instrRY} &
	$r_1 = r_2 = 0$ &
	\complabbrev &
	\issued{\instrY}, \completed{\instrRX} \\
	\complabbrev &
	\completed{\instrX}, \completed{\instrRY} &
	$x = y = 1$ &
	\complabbrev & 
	\completed{\instrY}, \completed{\instrRX} \\\hline
\end{tabular}
\end{center}

In this execution Processor A fetches and issues the two instructions in order, but the second instruction $\instrRY$ completes 
before the first. This may happen when the cache of Processor A already contains $y$, making the load of $y$ faster
than the \Store to $x$.
The same reasoning applies to the pipeline of Processor B.
Thus the \Stores of both processes complete after the \Loads, despite the memory operations of the \Stores being issued earlier.
This scenario leads to a final state with $x = y = 1$, 
but with both registers 0 ($r_1 = r_2 = 0$).

\end{multiexmp}

These ``reordering'' principles were extended by Intel's Pentium processors in the 1990s to include \emph{speculative execution}: 
while waiting for a branch condition to be resolved to true or false
(e.g., while waiting for a value to return from memory), a processor may
\emph{predict} the result and begin transiently executing down the corresponding path.
This transient execution may include executing memory operations associated with loads (but not stores)
out-of-order with instructions before the branch point.
If the prediction is later seen to be correct then the effects of the transient execution are committed; otherwise, the transient calculations are discarded and fetching restarts at the alternative (not-predicted) path. 
Where processor speeds are much faster than memory latency, speculative execution provides a substantial performance gain.
Such optimisations have led to significant security vulnerabilities, exposed by attacks such as Spectre~\cite{Spectre} and Meltdown~\cite{Meltdown}, which were only discovered decades after speculative execution and other optimisation mechanisms had become standard.


The out-of-order executions outlined above pose no issue on single-core processors, as they have been designed to preserve the sequential semantics of programs.
However, in the concurrent setting, these previously unobservable reorderings become potentially visible.
The crucial dependencies between instructions that modify or read local or shared variables can be summarised as follows.
\begin{definition}[Order dependence of variable accesses]
\labeldefn{order-deps}
If instruction $\aca$ is earlier in the pipeline than $\acb$, 
then $\acb$ can be completed before $\aca$ unless one of the following holds. \\
\begin{tabular}{rlr@{~}l}
	(i)& $\acb$ writes to a variable written to by $\aca$.
	&
	(ii)& $\acb$ writes to a variable read by $\aca$. 
	\\
	(iii)& $\acb$ reads a variable written to by $\aca$.\footnotemark 
	
	&
	(iv)& $\acb$ reads a \emph{shared} variable read by $\aca$.
\end{tabular}

\end{definition}
	\footnotetext{This may not always prevent reordering, due to forwarding (\refsect{forwarding}).} 
These checks are straightforward for a processor by examining the shared and local variables accessed by instructions in the pipeline.
The first three constraints are necessary to preserve sequential semantics on a single processor.
The last constraint is not necessary for the sequential semantics (and notably does not apply to accessing registers), but is required to ensure that each process observes the updates to a shared variable in a coherent global ordering.
For instance, consider the simple case of polling a (shared) variable $x$ that changes over time -- it is 
reasonable to assume that reads of $x$ in program order are intended to observe changes to $x$ chronologically. 
The four principles above still apply in the presence of fences and speculative execution, 
though other concerns -- such as preventing writes if there is an earlier unresolved branch -- also play a role.
Despite the difficulties instruction-level parallelism poses to formalising and reasoning about behaviours and potential security vulnerabilities,
these optimisations are likely to remain due to their associated computational efficiency.

\newcommand{\codefence}{\texttt{fence}}

\subsection{Fences and Memory Ordering Constraints}
\labelsect{fences}

Because memory operations can be reordered due to a range of factors, programmers need mechanisms to enforce the original order of instructions, if the logic of their algorithm so requires.
For example, in many data structures with fine-grained concurrency (\ie those that do not utilise locks for mutual exclusion), 
the precise ordering of memory operations is key to the correct interaction of competing processes.
To this end, every memory model provides \emph{fence} instructions types (also called \emph{barriers}),
and/or \emph{memory ordering constraints} (or \emph{access modes}) that can adorn existing instructions.

Many different types of fences exist; a \emph{full fence} ensures that all memory operations before it have been issued,
and prevents all later memory operations after it from executing early (see \refexmp{fence-tags}: x86 provides \texttt{mfence},
Arm provides \texttt{dmb}, and RISC-V provides \texttt{fence.rw.rw}).
Some processors come with more fine-grained fences; such a specialised fence may only affect specific types of memory operations 
(\Loads or \Stores), or may only affect instructions before or after it.
Memory ordering constraints may also be used to enforce ordering between otherwise independent instructions; 
the classic examples are the \emph{release} and \emph{acquire} constraints pioneered by 
Release Consistency \cite{ReleaseConsistency90} (see \refexmp{fence-tags-2} and \refsect{prelim-early}:
as an example, Arm provides both \texttt{LDAR} and \texttt{LDAPR} instructions for acquire loads).

\begin{multiexmp}

\exmpheading{Store buffering with fences}
\labelexmp{fence-tags}

\begin{wrapfigure}[4]{r}{0.25\textwidth}
  	\vskip -4mm
	\begin{tabular}{l||l}
		$x := 1$ ;		~&~ $y := 1$ ; \\
		\codefence ;	~&~ \codefence ; \\
		$r_1 := y$ ~&~ $r_2 := x$ \\
	\end{tabular}
\end{wrapfigure}

The store buffering litmus test (\refexmp{store-buffer}) can be modified to include fences.
Following \refexmp{sb-pipeline}, 
after fetching from the code base
	the pipeline of Process A becomes
[\plain{\instrX}, \codefence, $\instrRY$].

The fence instruction will not be issued while there are pending operations on main memory preceding it in the pipeline.
No memory operations after the fence will be issued until after the preceding fence has been processed.
Hence, if fences are inserted in the code of \emph{both} processes, a final state where $r_1 = r_2 = 0$ is no longer possible,
as the fences
enforce the originally specified ordering of memory operations.


\exmpsep

\exmpheading{Message passing with memory ordering constraints}
\labelexmp{fence-tags-2}

\begin{wrapfigure}[3]{r}{0.28\textwidth}
  	\vskip -4mm
	\begin{tabular}{l||l}
		$x := 1$ ;			~&~ $r_1 := \acqy$ ; \\
		$\rely := 1$	~&~ $r_2 := x$ \\
	\end{tabular}
\end{wrapfigure}

Recall the message-passing litmus test in \refexmp{mp}, with two of the instructions tagged with memory ordering constraints.
The \emph{release} constraint on the \Store to $y$ in Process A prevents the store from being issued prior to earlier instructions,
while the \emph{acquire} \Load of $y$ in Process B prevents later memory operations from being completed before its own.
As with the insertion of fences, this restores sequential order on the execution.

\end{multiexmp}

\subsection{Forwarding}
\labelsect{forwarding}

\emph{Forwarding} (or \emph{bypassing}) is a feature of processors
where a calculated value in the pipeline can be used by an instruction later in the pipeline.
It manifests in a program such as 
$
	x \asgn 1 \scomp r \asgn x
$.
According the principles of pipeline reorderings (\refsect{pipelining}), $r \asgn x$ must wait for $x \asgn 1$ to complete, as it is reading a variable being written by an earlier instruction.
Since the value being assigned to $x$ is already known, processors will typically simplify $r \asgn x$ to $r \asgn 1$
to avoid a potentially costly interaction with the main memory or local registers. 
After the simplification, the two instructions become independent,
and hence can be issued and completed in the reverse order while maintaining behaviour allowed by the sequential semantics.

\OMIT{
This may create apparent cycles in reasoning. Consider the following code.
\[
	(r \asgn x ; x \asgn 1 ; y \asgn x) \parallel (x \asgn y)
\]
Since $r \asgn x$ and $x \asgn 1$ are strictly ordered by dependence, it seems there is no way $r$ could read the value 1; however, due to forwarding,
$y \asgn x$ in the first process can proceed, which can result in $x$ being set to 1 by the second process, at which point the value 1 is read into $r$, 
and only then does the ``cause'', $x \asgn 1$ in the first process, finally take effect.

While on the surface this seems to destroy reasoning/coherence, the justification is mundane: the processor assumes that the programming/compiler is happy with $y \asgn x$
to be $y \asgn 1$ (always a possible outcome of the original code), and so simply treats it as such.  Then there is a value ``1'' in the system, and a separate update to $x$, 
and so it is not surprising that $r \asgn x$ can read that value - $y \asgn 1$ is an independent update.
}

\subsection{Multicopy Atomicity}\labelsect{multicopy-atomicity}
\labelsect{lmca}

The majority of modern CPU architectures (x86, Arm, and RISC-V) are said to possess the property of \emph{multicopy atomicity} -- every process observes the modifications to each variable in one consistent order.
For instance, suppose Process A modifies $x$ and then $y$.
On a multicopy-atomic system,
both Process B and Process C will observe those modifications in that order.
In contrast, on a system that \emph{lacks} multicopy atomicity, 
Processes B and C may disagree on the order of modifications, \ie
Process B may observe the modification to $x$ followed by the modification to $y$,
but Process C can simultaneously observe the modification to $y$ before the modification to $x$
(see \refexmp{iriw}).

\begin{exmp}[Independent reads of independent writes (IRIW)]
\labelexmp{iriw}

In this litmus test, let $(r_y := y_{(dep~r_x)})$ denote a \Load with an extra constraint that ``loading $y$ is dependent of the value of $r_x$''.%
\footnotemark

\begin{wrapfigure}[6]{R}{0.6\textwidth}
  \vskip -4mm
  \centering
  \begin{tabular}{l||l||c||c}
    $r_x := x$ ;           & $s_y := y$ ;           & $x := 1$ & $y := 1$ \\
    $r_y := y_{(dep~r_x)}$ & $s_x := x_{(dep~s_y)}$ & &
  \end{tabular}
  \\
  $\{(r_x = 1 \land r_y = 0) \land (s_x = 0 \land s_y = 1)\}$
  \\
  \begin{tabular}{lc}
  \SCmm, x86, Arm, RISC-V & \cross
  \\
  Power, C & \tick
  \end{tabular}
\end{wrapfigure}

The postcondition states that Process A observes $x \asgn 1$ before $y \asgn 1$, while Process B observes  them in the opposite order.
As Process A must first observe the change to $x$ and then observe that no change has yet occurred to $y$,
the postcondition must correspond to a global trace in the following form.
\[
	(x \asgn 1) \ldots (r_x \asgn x) \ldots (r_y \asgn y) \ldots (y \asgn 1)
\]
It is now impossible to insert the two \Loads of Process B into the trace in a consistent manner.
Therefore, this behaviour cannot be explained in terms of a single total ordering of the instructions.
However, this behaviour has been observed on Power architectures, and is allowed in \Clang.
\end{exmp}

\footnotetext{
This can typically be enforced by some notion of ``address dependency''. Such a dependency is weaker than fences, yet insists that the two instructions are ordered within the individual pipeline.
Fences cannot be used here, as they also affect the cache coherence system of Power, and thus prevent the IRIW behaviour. 
}

Currently, IBM's Power series is the only commercially available CPU architecture that lacks multicopy atomicity
(the NVIDIA PTX GPU also lacks multicopy atomicity \cite{nvidia-mm}).
This is due to its cache coherence system,
which effectively allows individual cores to communicate point-to-point via shared caches, before \Writes are flushed to the shared memory.

The lack of multicopy atomicity creates an extra level of complexity for describing memory models, as well as understanding their effects on code -- both at the assembler-level as well as in high-level languages to be compiled to the hardware.
To this end, Arm officially stipulates multicopy atomicity for their architectures from Version 8 in 2017 \cite{pulte-17}.
(In fact, Arm processors effectively have always been multicopy-atomic. Although multicopy atomicity was not explicitly stipulated by earlier versions of Arm, no manufacturer of Arm processors ever produced an implementation that lacked multicopy atomicity \cite{pulte-17}.)
Similarly, RISC-V insists on multicopy atomicity, and Intel's x86 architectures are also multicopy-atomic.

In practice, the complexity of dealing with an architecture that lacks multicopy atomicity leads programmers
to insert enough fences to restore multicopy atomicity (or indeed \SeqCst).
We therefore do not consider the lack of multicopy atomicity in this paper; 
a description of the behaviour of Power is given in \cite{ModellingARMv8},
and the difficulties of verification under Power's memory model are discussed in \cite{coughlin-22b}.

\OMIT{
\footnote{
While the \Clang memory model is not multicopy-atomic, 
when compiled to multicopy-atomic architectures it may be assumed to behave as if it is.
This is because compilers are unlikely to instrument the extra machinery required to exhibit these weaker behaviours,
i.e., to make it appear that updates to variables occur in different orders to different threads.
}
}


\section{The Development of Memory Models}\labelsect{prelim}

In this section, we sketch the history of how weak memory models have been described and addressed, and mention the development of significant hardware and software memory models. \reftable{memory-models-history} summarises the history with selected references. A reader interested purely in formalisms may skip this section.


\begin{table}[ht]
  \centering
  \caption{Selected history of hardware and software memory model formalisations}
  \labeltable{memory-models-history}
  \vspace{-4mm}
  \begin{tabular}{|c||c|c|c|c|c||c|c|}\hline
        & SPARC & Power & x86 & Arm & RISC-V & Java & C \\\hline\hline
  1990s & \cite{sindhu-91-tso} & \cite{IBM-POWER-93} & & & & & \\\hline
  2000s & & & \cite{owens-09}  & & & \cite{manson-05} & \cite{boehm-07} \\\hline
  2010s & & & & \cite{ModellingARMv8} & \cite{RISC-V} & & \cite{batty-11} \\\hline
  2020s & & & & \cite{alglave-21} & & & \\\hline
  \end{tabular}
\end{table}

\subsection{Early Memory Models}\labelsect{prelim-early}

\citet{LamportSC} made the first exploration on the notion of the ``intuitive correctness'' of concurrent programs in the 1970s, and formalised the concept of what later came to be known as \emph{\SeqCst}.
From the late 1980s onwards there were considerable efforts in defining the semantics of various instruction-level parallelism techniques, such as pipelining and out-of-order execution (see \refsect{history})~\cite{burch-dill-94,harman-07,jones-skakk-dill-02,shasha-snir-88}.
This saw the proposal of many theoretical memory models, aiming to categorise and unify observed and potential behaviours.
Examples include 
\emph{Weak Consistency}~\citep{dubois-86}, 
\emph{PRAM Consistency}~\citep{lipton-sandberg-88-pram}, 
\emph{Release Consistency} \citep{ReleaseConsistency90}, 
\emph{Cache Consistency}~\citep{goodman-91}, 
\emph{Processor Consistency}~\citep{goodman-91}, and 
\emph{Causal Consistency}~\citep{ahamad-95}.
These memory models were unified in 2004 by \citet{steinke-nutt-04}, who used a mathematical formalism to compare these memory models and to arrange them in a lattice. 
(In 2018, these theoretical models were again unified and compared \citep{senft-18-op,senft-18-temporal}.)

Before the advent of multicore processors, computers that had multiprocessing were mostly large-scale machines at scientific institutions, and only the programmers at these institutions needed to handle weak memory behaviours. Then, as multicore processors rose to the mainstream in the 2000s, weak memory behaviours became visible to more programmers in diverse areas. The focus of the research community thus shifted from defining new theoretical weak memory models to exploring the weak memory behaviour on multicore hardware and related higher-level programming languages.

\subsection{Hardware Memory Models}\labelsect{prelim-hard}

Many processor designers and vendors have attempted to codify and formalise the behaviours of their respective architectures, 
often with reference to the theoretical developments outlined in \refsect{prelim-early}. 
Compared to the memory models in \refsect{prelim-early}, which are mostly based on \Reads and \Writes only, the hardware memory models are more complex due to the specialised fences and atomic primitives of each architecture.
We consider several prominent examples below.

\paragraph{SPARC}
The Total Store Order (TSO) and Partial Store Order (PSO) memory models were defined for the SPARC architecture of Sun Microsystems~\cite{sindhu-91-tso}.
As the name suggests, TSO keeps the \Writes of each thread in program order, but a \Read that accesses some location $x$ can complete before an earlier \Write of the same thread to a location different from $x$.
PSO relaxes this further to maintain order only on \Writes to the same variable.
SPARC also defines a Relaxed Memory Order (RMO) model, 
which
``places no ordering constraints on memory references beyond those required for processor self-consistency'' 
\citep{SPARC-94}.
The TSO model of SPARC is formalised in Isabelle/HOL \cite{hou-21}.

\paragraph{x86}
The x86 family of architectures is widely implemented by various companies such as Intel and AMD. 
Aiming to devise a rigorous specification of the behaviour of x86, \citet{sarkar-09} developed the memory model x86-CC based on Causal Consistency (see \refsect{prelim-early}).
However, x86-CC was later found not to match the actual behaviour of the hardware in all cases.
Consequently, \citet{owens-09} developed x86-TSO, after observing similarities between the behaviour of x86 and the TSO memory model of SPARC.
To this date, x86-TSO is widely accepted as the correct description of the behaviour of x86 architectures. 
In~\cite{owens-09}, x86-TSO was introduced in terms of two formulations, one operational and one axiomatic. The operational formulation uses a write buffer to model the weak memory effects, and is influential on many later formalisations of memory models; we provide a version of it in \refsect{operat}.
The two key papers~\citep{owens-09,sarkar-09} that developed x86-TSO are summarised by \citet{sewell-10}.

\paragraph{Power} 
The Power family of processors
by IBM has a very relaxed memory model that lacks multicopy atomicity (see \refsect{multicopy-atomicity}), permitting many possible behaviours.
An axiomatic formulation for Power's memory model was first defined by \citet{IBM-POWER-93}, which was improved by \citet{adir-03} a decade later based on observed behaviours of litmus tests.
\citet{alglave-10,alglave-10-ext,alglave-14} also studied the Power memory model based on an axiomatic definition.
An alternative, operational approach to formalisation of Power was given by \citet{sarkar-11}.


\paragraph{Arm}
Whereas the desktop and laptop markets are dominated by x86, as of 2025 the mobile device market is dominated by Arm architectures.
Arm architectures previously did not forbid non-multicopy-atomic behaviours, but after a major revision in 2017, Arm has mandated multicopy atomicity from Version 8 onwards \cite{pulte-17,alglave-21}.
There have been numerous works on specifying versions of the Arm memory model using different formalisms, such as \cite{ModellingARMv8}. The development history is included in the latest official paper on the Arm memory model \cite{alglave-21}.

\paragraph{RISC-V}
The RISC-V architecture~\citep{RISC-V}, first released in 2010, is an open-source architecture that attempts to remove some of the problems associated with commercial-in-confidence manufacturing. Its memory model is based on experiences at Arm, and includes many recent developments in weak memory theory, including support for release-acquire primitives.

\paragraph{Specialised Hardware}
Recent research effort has focussed on GPUs \cite{alglave-15-gpu,sorensen-16-gpu,lustig-19} and FPGAs \cite{iorga-21},
and their combination with CPUs \cite{CompoundMMs-23}.
GPUs in particular focus on significant parallelisation of processing power, and hence would benefit from formalisation; NVIDIA has provided a rigorous description of its memory model
\cite{nvidia-mm}, which lacks multicopy atomicity (see \refsect{lmca}).  Earlier formalisations determined a type of total-store-order model \cite{alglave-15-gpu}, while more recent formalisations suggest a weaker model again \cite{lustig-19}.

\paragraph{Earlier Hardware Memory Models}
Intel's \emph{Itanium} processors \cite{abe-17} never became as commercially popular as other processors,
and have since ceased production. 
Similarly, DEC's \emph{Alpha} processors were prominent for a time, but have also ceased production.

\subsection{Software Memory Models}\labelsect{prelim-soft}

In the interest of portability, most languages provide their own abstraction and generalisation of the underlying hardware weak memory models, so that programmers need only consider a single set of primitives. 
This also constrains compilers to certain regular transformations in the presence of instruction dependencies.
We consider some of the major languages below with a focus on the development of formal models; 
we recommend the wider survey of software memory models by \citet{mois-21} for those interested in a comparison of features and limitations.

\paragraph{Java.}
Sun Microsystems released the Java programming language in 1996, with multi-threading and shared-memory behaviour informally described in~\citep{java-spec-96}.

This was immediately followed by multiple independent efforts to formalise the memory model of 
Java~\citep{attali-98,gont-98,higham-98}.  
There is a formulation that explicitly used local working memories~\cite{boerge-99}, and a more abstract formulation with event structures~\cite{cenc-99}.
In 1999, however, the original prose memory model turned out to contain flaws~\citep{pugh-99,pugh-00}. 
Subsequent work~\citep{arvind-00,manson-01,roych-02a,yang-01,yang-02,awhad-03,manson-04} attempted to revise the Java memory model, culminating in~\cite{manson-05}, which describes the new Java memory model as part of Java 5.0.

After the formulation in~\cite{manson-05} became accepted as the official Java model, a few alternative specifications continued to emerge~\cite{yang-05-umm,jagad-10,lochb-13,bender-19}. On the compiler-related side, \citet{sevcik-08} identified common optimisations performed Java compilers that do not comply with the Java memory model.


\paragraph{C and C++.}
Unlike Java, which included concurrency constructs in its specification from the very start, the C and C++ programming languages were originally designed without built-in mechanisms for concurrent programming. Instead, concurrency was enabled by a separate library of multi-threading constructs. However, as \citet{boehm-05} pointed out in 2005, ``threads cannot be implemented as a library'': the correctness of concurrent code cannot be guaranteed when the concurrency mechanisms are designed separately from the core language and its compiler.

Boehm and others worked as part of the International Organisation for Standardisation (ISO) on the formal semantics of concurrent C and C++ programs~\citep{boehm-07,boehm-08}. The resulting 2011 ISO standards of the C and C++ programming languages were the first time that multi-threading in both languages was formally defined, and the memory model defined by these ISO standards is commonly referred to as the \emph{C memory model} (or \emph{C11} in the literature, owing to its inception in 2011).
The definition of the C memory model was then consolidated into a formal axiomatic formulation by \citet{batty-11}, and later into a formal operational formulation by \citet{nienhuis-16}.

The prominence of the \Clang language and its memory model has drawn much attention in the context of verification, and many recent publications are devoted to program logics~\citep{wright-21-og}, model checkers~\citep{kokolo-18}, or other problems~\citep{singh-21,singh-22,colvin-22}; more of these are covered in \refsect{tools}.

Software memory models are often subject to more potential issues, due to the wider range of transformations open to a compiler, on top of the weak memory behaviours on the target hardware.
These issues include the ``out of thin air'' (see \refexmp{oota}) and ``read from untaken branch'' behaviours.
These have been particularly addressed in the context of \Clang,
and the question about how to forbid out-of-thin air (and similar) behaviours without overly constraining standard
compiler optimisations is a subject of ongoing debate in the standardisation committees and the academic 
community~\cite{boehm-14,batty-15,jeffrey-16,pichon-16,ou-demsky-18,chakra-19}.
The most recent \Clang reference explicitly gives a version of out-of-thin-air behaviour, stating that it is a
theoretically allowed behaviour, but with a caveat that no compiler should ever implement it.

\begin{exmp}[LB+ctrls]
\labelexmp{oota}

\begin{wrapfigure}[7]{r}{0.42\textwidth}
	\vskip -2mm
	\begin{tabular}{c}
		\begin{tabular}{l||l}
			$r_y := y$ ;          & $r_x := x$ ;  \\
			$\If r_y = 1 \Then$ & $\If r_x = 1 \Then$ \\
			\quad $x \asgn 1$ & \quad $y \asgn 1$ \\
		\end{tabular}
		\\
		$\{ x = 1 \land y = 1 \}$ \\
		\begin{tabular}{ll}
		\SCmm, x86, Arm, RISC-V & \cross \\
		\Clang & \tick (?)
		\end{tabular}
	\end{tabular}
\end{wrapfigure}

This infamous litmus test,
potentially exhibiting an
``out of thin air'' behaviour,
considers a final state where $x=1$ and $y=1$, which is not reachable under any standard interpretation of program
language semantics, whether weak memory concerns are addressed or not. 
For Process A to execute $x \asgn 1$, it must have read the value 1 for $y$ (via $r_y \asgn y$).
This can only happen if the branch condition of Process B evaluates to true, i.e., if Process B previously read the
value 1 for $x$.  Since this requires circular reasoning, it must be the case that both $x$ and $y$ are 0 at the end of
execution on any
hardware. 
However, outright banning the behaviour at the software level is problematic because
it is unclear how to prevent this behaviour without also preventing otherwise sensible 
optimisations in other situations.
\end{exmp}

With the memory model of Power lacking multicopy atomicity and hence being harder to handle (see \refsect{prelim-hard}), there have been works that focus on the compilation from C to Power while respecting the memory models of both ends~\cite{batty-12,sarkar-12,lahav-17},
and more general compiler optimisations under the C memory model~\cite{podkopaev-19,dodds-18,geeson-compiler-24}.

\paragraph{The Linux Kernel.}
Alglave et al. \cite{alglave-18} 
provide a formalisation of the memory model of the Linux kernel.
This model is to further codify the complex interaction of the wide variety of hardware memory models and 
the C language memory model, supporting some of the common patterns specific to Linux kernel programming.
The formalisation involved interaction with the developer community, and the effort helped identify and resolve 
several issues, especially regarding locks and the read-copy-update (RCU) paradigm.
As such this work represents a strong argument for the practicality of involving formalisms in non-trivial software development.

\paragraph{Other Programming Languages.}
Every programming language that supports shared-memory concurrency has interplay with weak memory models. For programming languages such as
JavaScript~\citep{ecmascript-20,javascript-20}, 
OCaml~\citep{dolan-18}, 
OpenCL~\citep{OpenCL} and
Rust~\citep{rust-lang},
there have been works that study their own memory models, or how they are affected by the weak memory models of the languages to which they may be compiled.
A potential simplification for this effort for many languages is via a memory model for an intermediate language such as
LLVM
\citep{chakra-vaf-17-llvm,lee-23}, and through the investigation of general properties related to safe optimisation
strategies~\cite{gopal-23}.


%% file: mechanistic.tex
\section{Introduction to Operational Formalisms}\labelsect{operat}
\labelsect{mechanistic}

\newcommand{\cat}{\mathbin{\raise 0.8ex\hbox{\scriptsize{$\frown$}}}}

\newcommand{\labeloprule}[1]{\text{(#1)}}

\newcommand\refoprule[1]{\text{(#1)}}

\newcommand{\eval}[2]{\mathsf{eval}_{#1}(#2)}
\newcommand{\evalse}{\eval{\regs}{e}}
\newcommand{\buf}{\textsf{b}}
\newcommand{\cmd}{\textsf{c}}
\newcommand{\regs}{\textsf{s}}
\newcommand{\sdef}{\mathrel{\widehat=}}

\newcommand{\silent}{\tau}

\newcommand{\core}{\mathtt{core}}
\newcommand{\lcfg}{\core}
\newcommand{\labelrule}[1]{\label{eq:#1}}
\newcommand{\refrule}[1]{\ref{eq:#1}}

\newcommand{\csep}{~~|~~}

\newcommand{\ttdef}{\mathrel{::=}}
\newcommand{\attdef}{&\ttdef&}
\newcommand{\asdef}{&\sdef&}

\newcommand{\fun}{\rightarrow}

\newcommand{\double}[2]{\langle #1 , #2 \rangle}
\newcommand{\triple}[3]{\langle #1 , #2 , #3 \rangle}

\newcommand{\ebuf}{[~]}

\newcommand{\cmdregsinline}[2]{\double{#1}{#2}}
\newcommand{\cmdregs}[2]{\left\langle\!\!\begin{array}{c}#2 \\ #1 \end{array}\!\!\right\rangle}
\newcommand{\cmdregsc}[1]{\cmdregs{\cmd}{#1}}
\newcommand{\cmdregscs}{\cmdregsc{\regs}}
\newcommand{\cmdregscsp}{\cmdregs{\cmd'}{\regs'}}
\newcommand{\cmdregscsi}[1]{\cmdregs{\cmd_{#1}}{\regs_{#1}}}
\newcommand{\cmdregscspi}[1]{\cmdregs{\cmd'_{#1}}{\regs'_{#1}}}
\newcommand{\bufcoreinline}[2]{#1~\circ #2}

\newcommand{\overlay}[2]{\ooalign{\hss{#1}\hss\cr{#2}}}

\newcommand{\bufcore}[2]{\raisebox{-4pt}{$\begin{array}[c]{c}#1 \\ \overbracket{~#2~\vphantom{'}} \end{array}$}}
\newcommand{\bufcoreb}[1]{\bufcore{\buf}{#1}}
\newcommand{\bufcorebcs}{\bufcoreb{\cmdregscs}}
\newcommand{\bufcorebcsi}[1]{\bufcore{\buf_{#1}}{\cmdregscsi{#1}}}

\newcommand{\bufcorebc}{\bufcore{\buf}{\lcfg}}
\newcommand{\bufcorebcp}{\bufcore{\buf}{\lcfg'}}

\newcommand{\Types}[1]{\mathbf{#1}}

\newcommand{\Pid}{\Types{Pid}}
\newcommand{\Val}{\Types{Val}}
\newcommand{\Expr}{\Types{Expr}}
\renewcommand{\LVar}{\Types{LVar}}
\newcommand{\SVar}{\Types{SVar}}
\newcommand{\Cmd}{\Types{Cmd}}
\newcommand{\Skip}{\mathbf{skip}}

\newcommand{\Mem}{\Types{Mem}}
\newcommand{\LocS}{\Types{LocState}}
\newcommand{\LocC}{\Types{LocConfig}}
\newcommand{\ParC}{\Types{ParConfig}}
\newcommand{\Syst}{\Types{Syst}}

\newcommand{\Buf}{\Types{Buf}}

\newcommand{\Tid}{\mathbf{Tid}}
\newcommand{\Prog}{\mathbf{Prog}}
\newcommand{\TState}{\mathbf{TState}}
\newcommand{\MState}{\mathbf{MState}}
\newcommand{\ActualMem}{\mathbf{ActualMem}}
\newcommand{\BufState}{\mathbf{BufState}}

\newcommand{\Memory}{\mathtt{mem}}
\newcommand{\multicores}{\mathtt{mcores}}
\newcommand{\mainmem}[2]{\begin{array}{c} #1 \\ \overbrace{~#2~\vphantom{'}} \end{array}}


Before giving a general survey of the field
we provide an example formalisation:
a simplification of
the formalisation of x86 given by the influential work by Owens et al. \cite{owens-09,sewell-10}.
This formalisation is \emph{operational}, meaning described by way of individual steps of the system,
and as such is often easier to follow than other types of semantics \cite{Jones-18}.
We also choose it as it is based around a microarchitectural feature, and as such provides a contrast to 
\emph{axiomatic} descriptions, which purposefully abstract away from such considerations.
In this formalisation,
weak memory effects (specifically, loads appearing to take effect before earlier stores) emerge
by the use of a \emph{write buffer} in each core (acting similarly to a pipeline (\refsect{pipelines})
or as part of the cache system \cite{mckenney2010memory}). 
We first give an operational semantics for a standard sequential language, as if instructions were executed in
the order specified by the program text; after extending that semantics with extra rules covering write buffers 
the weaker behaviours of x86 (Total Store Order) become apparent.
More complex formalisations based on microarchitectural mechanisms
are outlined in \refsect{mechanistic-further}; see also \refsect{mechanistic-survey}.



\newcommand{\Update}[3]{#1[#2 \asgn #3]}

\newcommand{\List}[1]{\textit{List}~#1}

\paragraph{Notation.}
Let $f:A\fun B$ be a function from domain type $A$ to range type $B$. 
Viewing a function as a set of pairs, we sometime write $a \mapsto b$ in place of $(a,b)$ when $f(a) = b$.
Also, we write $\Update{f}{a}{b}$ to denote the updated function that maps $a$ to $b$ and maps every other $x\in A$ to its original $f(x)$.
A typical list is written as $[a_1,a_2,\ldots a_n]$, and the empty list is written as $\ebuf$. 
We use `$\cat$' to append or prepend elements and lists,
for example, $l \cat m$ is the list formed by appending lists $l$ and $m$,
and if $a_1, a_2$ and $a_3$ are elements,
$ a_1 \cat [a_2,a_3] = [a_1,a_2,a_3] = [a_1,a_2] \cat a_3$.

\subsection{Syntax}\labelsect{operat-basics}


We begin by introducing an assembly-like programming language.
The basic types include
\emph{values} $\Val$, 
\emph{local variables} $\LVar$,
and
\emph{shared variables} $\SVar$.
By convention,
$v$ ranges over $\Val$,
$r$ ranges over $\LVar$ (registers), and
$x,y$ range over $\SVar$.
An \emph{expression} $e$ can be formed from any of the standard arithmetic, logical operators, and local variables. 

\newcommand{\ffence}{\mathsf{fence}}

The syntax of \emph{commands} $\Cmd$ is defined as follows. 
\begin{equation}\labeleqn{defn-syntax}
  c ::=
      r := e 
      ~|~ x := e 
      ~|~ r := x 
      ~|~ \ffence
      ~|~ c_1 ; c_2 
  	  ~|~ \Skip 
\OMIT{
  c ::=~& \Skip \\
      |~& r := e &\text{(local assignment)}\\
      |~& x := e &\text{(writing to shared memory)}\\
      |~& r := x &\text{(reading from shared memory)}\\
      |~& c_1 ; c_2 &\text{(sequential composition)} 
}
\end{equation}
A command $c$ may be one of three ``update'' instructions:
	  a local assignment $(r \asgn e)$,
	  a \Store to shared memory $(x \asgn e)$,
	  or a \Load from shared memory $(r \asgn x)$.
In all cases we assume expression $e$ refers only to registers and values.
These instructions correspond to different uses of the x86 instruction type \texttt{mov};
for instance,
``\texttt{mov r1,r2}'' ($r_1 \asgn r_2$),
``\texttt{mov [rx],r}'' ($x \asgn r$),
and
``\texttt{mov r,[rx]}'' ($r \asgn x$),
where register \texttt{rx} has the value of a location in memory (abstractly corresponding to a shared variable $x$).
In addition to the three ``update'' instructions, a command could be a $\ffence$ (\eg \texttt{mfence} in x86).
Instructions can also be composed sequentially, $c_1 \scomp c_2$
(though this does not guarantee sequential execution, of course), 
and a
terminated command is $\Skip$.

A \emph{multicore system} is formed from a single \emph{shared state} of type $\SVar \fun \Val$,
and two or more individual \emph{cores}, composed in parallel ($\pl$).
Each core is itself formed from a \emph{local state} of type $\LVar \fun \Val$ and a command of type $\Cmd$.

\OMIT{
A \emph{memory state} $\Mem : \SVar \fun \Val$ assigns a value to each shared variable.
Each process $i$ runs a command in some \emph{local state} $\LocS_i: \LVar_i \fun \Val$, which assigns a value to each local variable in $\LVar_i$.
The pairing of a command and a local state is called a \emph{local configuration} $\LocC_i : \Cmd \times \LocS_i$,
which we write as $\cmdregs{c}{s_i}$, where $c \in \Cmd$ and $s_i \in \LocS_i$.
Processes are composed in parallel to form a \emph{parallel configuration} 
\[ \ParC : \LocC_1 \times \cdots \times \LocC_n ,\] 
which we write as $\cmdregscsi{1} \pl \ldots \pl \cmdregscsi{n}$.

Globally, a system consists of a shared memory and a set of processes running in parallel. Hence, \emph{system states} have type $\Syst : \Mem \times \ParC$,
which we write as 
$\mainmem{\Memory}{\multicores}$, where $\Memory$ is the memory and $\multicores$ is the parallel system.
}


For instance, recall the store-buffer (\SB) litmus test from \refsect{prelim-litmus}.
The initial configuration for this system \refeqn{initial-config-SB} is that shared variables $x$ and $y$ map to 0 in the global state,
and there are exactly two cores,
each of which has a single local variable mapping to 0 and containing the corresponding code.
To aid readability we write the structures in vertical form.

\begin{equation}
\labeleqn{initial-config-SB}
\mainmem{
     \{ x \mapsto 0 ~,~ y \mapsto 0 \}
}{
	\cmdregs{
     x:=1 \scomp r_1:= y
	 }{
     \{ r_1 \mapsto 0 \}
	 }
	 \pl 
	 \cmdregs{
	 y:=1 \scomp r_2:= y
	 }{
     \{ r_2 \mapsto 0 \} 
	 }
}
\end{equation}

\subsection{Semantics}\labelsect{operat-transitions}

Each possible execution of the concurrent system is viewed as a sequence of \emph{memory events},
which may be a \Read-event $\ReadEvent{}{x}{v}$, a \Write-event $\WriteEvent{}{x}{v}$, a $\ffence$-event, or a silent event $\silent$.
A \Read $\Revxv$ observes the value $v$ for $x$ from shared memory,
whereas a \Write $\Wevxv$ updates the variable $x$ to value $v$ in shared memory.
A $\ffence$-event is issued by a $\ffence$-instruction.
A silent event $\silent$ denotes some register-only instruction that is resolved locally. 

\begin{equation}
\labeleqn{defn-mem-events}
\lbl \ttdef 
\ReadEvent{}{x}{v} \csep \WriteEvent{}{x}{v} \csep \ffence \csep \silent
\end{equation}

\emph{Local Operational Rules.} 
The behaviour of a process is defined by transitions of the form 
$\core \tra{\lbl} \core'$,
where a $\core$ is formed from a local state $s$ and command $\cmd$,
and $\lbl$ is the generated memory event.
\reffig{rules} gives the transition rules for the five command types in \refeqn{defn-syntax}.

\begin{figure}

\begin{eqnarray*}
\hline
\end{eqnarray*}

\begin{center}
Basic commands
\end{center}

\begin{gather*}
  \labeloprule{seq}~
  \Rule{\cmdregs{ c_1}{ s} \tra{\lbl} \cmdregs{c_1' }{ s'}}
       {\cmdregs{ c_1 \scomp c_2}{ s} \tra{\lbl} \cmdregs{c_1' \scomp c_2} { s'}}
  \quad 
  \labeloprule{skip}~
  \Rule{}{\cmdregs{ \Skip \scomp c }{ s} \tra{\silent} \cmdregs{ c }{ s }}
\end{gather*}

\begin{gather*}\label{eq:loc}
  \begin{array}{l}
  \labeloprule{local}\quad
  \Rule{}{\cmdregs{ r:=e }{ s } \tra{\silent} \cmdregs{ \Skip }{ \Update{s}{r}{\evalse} }}
\\
  \labeloprule{load}\quad
  \Rule{}{\cmdregs{r:=x}{s} \ttra{\Revxv}  \cmdregs{\Skip }{ \Update{s}{r}{v}}}
  \end{array}
  \begin{array}{l}
  \labeloprule{store}\quad
  \Rule{}{\cmdregs{ x:=e }{ s} \ttra{\WriteEvent{}{x}{\evalse}} \cmdregs{\Skip }{ s}}
  \\
  \labeloprule{fence}\quad
  \Rule{}{\cmdregs{ \ffence }{ s } \ttra{\ffence} \cmdregs{ \Skip }{ s }}
  \end{array}
  ~\\
  ~\\
  \hline
\end{gather*}



\begin{center}
System level
\end{center}

\begin{equation*}\label{eq:par-config}
  \labeloprule{par-config}\quad
  \Rule{\core_i \tra{\ell} \core'_i}
       {\core_1 \pl \ldots \pl \core_i \pl \ldots \pl \core_n
        ~~\tra{\lbl}~~
        \core_1 \pl \ldots \pl \core'_i \pl \ldots \pl \core_n}
\end{equation*}

\begin{gather*}
  \labeloprule{mem-read}
  ~~
  \Rule{\Memory(x)=v \\ \multicores \ttra{\Revxv} \multicores'}
       {\mainmem{\Memory}{\multicores}
        \ttra{\Revxv} 
        \mainmem{\Memory}{\multicores'}}
  \quad
  \labeloprule{mem-write}
  ~~
  \Rule{\multicores \ttra{\Wevxv} \multicores'}
       {\mainmem{\Memory}{\multicores}
        \ttra{\Wevxv} 
        \mainmem{\Update{\Memory}{x}{v}}{\multicores'}}
\\
  \labeloprule{mem-$\silent$}
  ~~
  \Rule{\multicores \tra{\silent} \multicores'}
       {\mainmem{\Memory}{\multicores}
        \tra{\silent}
        \mainmem{\Memory}{\multicores'}}
	\\
	~\\
	\hline
\end{gather*}

\caption{Operational rules for a sequentially consistent multicore system}
\labelfig{rules}
\Description{TODO}

\end{figure}

\refoprule{seq} states that a sequential composition of commands $c_1 \scomp c_2$ may always take a step of $c_1$.
If $c_1$ has terminated (\ie is $\Skip$), then $c_2$ may be executed via \refoprule{skip}.

In \refoprule{local}, a local assignment $(r \asgn e)$ 
updates $r$ in the local state to $\evalse$, the value of expression $e$ evaluated at the local state $s$;
since it is resolved purely in the local configuration, 
it is a silent transition, having no interaction with shared memory.

In \refoprule{load}, $r \asgn x$ 
returns some value $v$ for $x$ from shared memory, and updates the register $r$ accordingly.
Similarly, in \refoprule{store}, $x \asgn e$ 
writes the (locally evaluated) value $\evalse$ to $x$ in shared memory.
The final $\Skip$ in each rule indicates that the instruction has finished executing,
and will eventually be eliminated via \refoprule{skip}.

A $\ffence$-instruction -- which has no effect in a sequentially consistent system -- simply generates a $\ffence$-event; its influence will be seen in the next section on write buffers.

\emph{System-Level Operational Rules.} 
In \refoprule{par-config},
a system with $n$ cores operating in parallel may take a step when any 
single core ($i$) takes a step.
At the topmost level these generated memory events interact with main memory, $\Memory$.
In these rules, 
$\multicores$ is the parallel composition of processes, as in \refoprule{par-config}.
In \refoprule{mem-read}, a \Read-instruction of $x$, where the value of $x$ is $v$ in shared memory, 
may proceed, and the shared memory remains unchanged --
the precondition for this rule prevents incorrect values being read in \refoprule{load}.
In \refoprule{mem-write}, if any process $i$ issues a \Write of the value $v$ to the shared variable $x$ then the memory is updated accordingly.
Finally, \refoprule{mem-$\silent$} states that silent events have no effect on main memory; a similar rule holds for $\ffence$
instructions.

\emph{Termination.}\labelsect{operat-sc}
A process is said to \emph{terminate} when it reaches a single $\Skip$ command.
A system is said to \emph{terminate} if all processes terminate,
and a system state, $s$, is said to be \emph{reachable} if the system from some starting state can terminate in $s$ via a sequence of top-level transitions 
of \refoprule{mem-read}, \refoprule{mem-write}, and \refoprule{mem-$\silent$}.



\begin{exmp}[\SB reachable state under \SeqCst]
\labelexmp{op-sc}

\begin{wrapfigure}[11]{r}{0.56\textwidth}
\begin{minipage}{0.56\textwidth}
\vskip -4mm
\begin{eqnarray*}
	&&
		\cmdregs{
		 x \asgn 1 \scomp r_1 \asgn y
		 }{
		 \{ r_1 \mapsto 0 \}
		 }
		 \pl 
		 \cmdregs{
		 y \asgn 1 \scomp r_2 \asgn y
		 }{
		 \{ r_2 \mapsto 0 \} 
		 }
	\\
	&\ttra{\Wev{x}{1}}&
		\cmdregs{
		 r_1 \asgn y
		 }{
		 \{ r_1 \mapsto 0 \}
		 }
		 \pl 
		 \cmdregs{
		 y \asgn 1 \scomp r_2 \asgn y
		 }{
		 \{ r_2 \mapsto 0 \} 
		 }
	\\
	&\ttra{\Rev{y}{0}}&
		\cmdregs{
		 \Skip
		 }{
		 \{ r_1 \mapsto 0 \}
		 }
		 \pl 
		 \cmdregs{
		 y \asgn 1 \scomp r_2 \asgn y
		 }{
		 \{ r_2 \mapsto 0 \} 
		 }
	\\
	&\ttra{\Wev{y}{1}}&
		\cmdregs{
		 \Skip
		 }{
		 \{ r_1 \mapsto 0 \}
		 }
		 \pl 
		 \cmdregs{
		 r_2 \asgn y
		 }{
		 \{ r_2 \mapsto 0 \} 
		 }
	\\
	&\ttra{\Rev{x}{1}}&
		\cmdregs{
		 \Skip
		 }{
		 \{ r_1 \mapsto 0 \}
		 }
		 \pl 
		 \cmdregs{
		 \Skip
		 }{
		 \{ r_2 \mapsto 1 \} 
		 }
\end{eqnarray*}
\end{minipage}
\end{wrapfigure}

Recall the store buffering litmus test from \refexmp{store-buffer}.
To show that the state $x = 1 \land y = 1 \land r_1 = 0 \land r_2 = 1$ is reachable under SC
from the start state 
it suffices to construct a sequence of transitions that leads to the following terminated system state.
~\\
$
\mainmem{
     \{ x \mapsto 1 , y \mapsto 1 \}
}{
	\cmdregs{
     \Skip
	 }{
     \{ r_1 \mapsto 0 \}
	 }
	 \pl 
	 \cmdregs{
	 \Skip
	 }{
     \{ r_2 \mapsto 1 \} 
	 }
}
$ \\
As shown to the right
the system may evolve by executing both instructions of Core A (left-hand side), followed by both instructions of Core B (right-hand side).
The instructions of both cores are interleaved by \refoprule{par-config}.
At each stage, the local state is updated, and so is the shared memory.
For brevity, we omit steps eliminating leading $\Skip$s by \refoprule{skip}, and
omit main memory.
The sequence of memory operations is consistent with the memory transitions \refoprule{mem-read} and \refoprule{mem-write},
and hence this is a valid sequence of steps, resulting in the final state.

\end{exmp}

\subsection{Write Buffers for x86}\labelsect{operat-tso}

We now augment the system
of \refsects{operat-basics}{operat-transitions} with an explicit write buffer for each core,
following~\cite{owens-09}.
%
A \emph{write buffer}, of type $\Buf$, is a list of pairs $(x,v)$, where $x$ is a shared variable and
$v$ is a value being written to $x$, 
representing
a \emph{pending} \Write of $v$ to $x$. 
The head (left-hand side) of the write buffer is the oldest \Write, with new \Writes being appended to the end (right-hand side) of the write buffer.
A core now contains a write buffer, in addition to the local state and command;
the system-level structure is otherwise unchanged.


\begin{wrapfigure}[6]{r}{0.52\textwidth}
\vskip -4mm
$
\mainmem{
     \{ x \mapsto 0 ~,~ y \mapsto 0 \}
}{
	\bufcore{\ebuf}{
	\cmdregs{
     x:=1 \scomp r_1:= y
	 }{
     \{ r_1 \mapsto 0 \}
	 }
	 }
	 \pl 
	 \bufcore{\ebuf}{
	 \cmdregs{
	 y:=1 \scomp r_2:= y
	 }{
     \{ r_2 \mapsto 0 \} 
	 }
	 }
}
$
\end{wrapfigure}
\noindent
For instance,
the starting configuration for \SB is similar to that 
of \refeqn{initial-config-SB},
except that each core also has an (initially empty) buffer.




\reffig{rules-buffers} gives the operational rules of a $\core$ (formed from a local state $s$ and command $\cmd$) interacting with its local write buffer $\buf$.
All previous transitions in \reffig{rules} still apply as-is,
with the combination of \refoprule{x86-write-buf} and \refoprule{x86-read-mem} allowing \Loads to take effect before \Stores.

\begin{figure}

\begin{gather*}
  \text{(x86-write-buf)} ~~
  \Rule{\lcfg \ttra{\WriteEvent{}{x}{v}} \lcfg'}
       {\bufcorebc \tra{\silent} \bufcore{\buf \cat (x,v)}{\lcfg'}}
  \labelrule{tso-2} 
  \quad
  \text{(x86-write-mem)} ~~
  \Rule{}
       {\bufcore{(x,v) \cat \buf}{\lcfg} \ttra{\WriteEvent{}{x}{v}} \bufcorebc}
  \\
  \text{(x86-read-mem)} ~~
  \Rule{
  	\lcfg \ttra{\Revxv} \lcfg'
  	\\
  	\lnot\exists (x,\_) \in \buf
  }{
    \bufcorebc 
  		\ttra{\Revxv} 
    \bufcorebcp
  }
  \quad
  \labelrule{tso-3}
  \text{(x86-read-buf)} ~~
  \Rule{
  	\lcfg \ttra{\Revxv} \lcfg'
  	\\
  	\text{$(x,v)$ is the rightmost $(x,\_)$ in $\buf$}
  }{
    \bufcorebc \tra{\silent}
    \bufcorebcp
  }
  \\
  \labelrule{tso-0} 
  \text{(x86-$\silent$)} ~~
  \Rule{\lcfg \tra{\silent} \lcfg'}
       {\bufcorebc \tra{\silent} \bufcorebcp}
\qquad \qquad
  \text{(x86-fence)} ~~
  \Rule{\lcfg \ttra{\ffence} \lcfg'}
       {\bufcore{\ebuf}{\lcfg} \tra{\silent} \bufcore{\ebuf}{\lcfg'}}
  \\
\end{gather*}

\vspace{-6mm}
\caption{Operational rules for write buffers}
\labelfig{rules-buffers}
\Description{TODO}

\end{figure}

\begin{multiexmp}

\exmpheading{\SB with write buffers}
\labelexmp{operat-tso-1}

\begin{wrapfigure}[4]{r}{0.52\textwidth}
\vskip -6mm
Core A \\
$
	\bufcore{\ebuf}{\cmdregs{x \asgn 1 \scomp r_1 \asgn y}{\{r_1 \mapsto 0\}}}
	\tra{\silent}
	\bufcore{[(x,1)]}{\cmdregs{\Skip \scomp r_1 \asgn y}{\{r_1 \mapsto 0\}}}
$
\end{wrapfigure}
Recall the litmus test from \refexmp{store-buffer} and \refexmp{op-sc}.
We demonstrate an execution of \SB under TSO that leads to a state where $r_1=0 \land r_2=0$ 
-- unreachable under \SeqCst --
starting from the usual global configuration where the shared variables are 0 and the buffers are empty.
To aid readability, we annotate each event with A or B, indicating the core from which it was issued.
Suppose Core A executes first;
executing the first instruction $(x \asgn 1)$ adds a $\Write$ to the write buffer by \refoprule{x86-write-buf}.

\begin{wrapfigure}[5]{r}{0.62\textwidth}
\vskip -2mm
Core A \\
$
\ldots
	\ttra{\ReadEvent{A}{y}{0}}
	\bufcore{[(x,1)]}{\cmdregs{\Skip}{\{r_1 \mapsto 0\}}}
	\ttra{\WriteEvent{A}{x}{1}}
	\bufcore{\ebuf}{\cmdregs{\Skip}{\{r_1 \mapsto 0\}}}
$
\end{wrapfigure}
Then by \refoprule{x86-read-mem}, Core A continues by reading $y$ directly from the shared memory, as there are no pending \Writes to $y$ in its write buffer; as such, this \Read occurs \emph{before} the \Write to $x$ (which is not possible under \SeqCst).
Finally, the \Write in the write buffer is sent to the shared memory according to (x86-write-mem). 

\begin{wrapfigure}[5]{r}{0.68\textwidth}
\vskip -2mm
Core B \\
$
	\bufcore{\ebuf}{\cmdregs{y \asgn 1 \scomp r_2 \asgn x}{\{r_2 \mapsto 0\}}}
	\ttra{\ReadEvent{B}{x}{0}}
	\ldots
	\ttra{\WriteEvent{B}{y}{1}}
	\bufcore{\ebuf}{\cmdregs{\Skip}{\{r_2 \mapsto 0\}}}
$
\end{wrapfigure}
Using similar reasoning Core B reads $x = 0$ and then stores $y \asgn 1$.

\quad
Interleaving the steps of each local configuration may give the operation sequence
$
[
	\ReadEvent{A}{y}{0},
	\ReadEvent{B}{x}{0},
	\WriteEvent{A}{x}{1},
	\WriteEvent{B}{y}{1}
]
$.
Combining this sequence of transitions with the shared memory, we arrive at a TSO system state where $x=y=1 \land r_1=r_2=0$
--- 
an unreachable state under \SeqCst.

\exmpsep

\exmpheading{Forwarding}
\labelexmp{operat-bypassing}

\begin{wrapfigure}[12]{R}{0.50\textwidth}
\begin{minipage}{0.50\textwidth}
\vskip -7mm
\begin{equation*}
\begin{array}{r@{~}l}
	& \bufcore{\ebuf}{
		\cmdregs{x \asgn 1 \scomp y \asgn 2 \scomp z \asgn 3 \scomp r \asgn y}
		          {\{r \mapsto 0\}}
		}
	\\
	\tra{\silent}
	&
	\ldots
	\tra{\silent}
	\ldots
	\\
	\tra{\silent}
	& \bufcore{[(x,1), (y,2), (z,3)]}{
		\cmdregs{r \asgn y}
	            {\{r \mapsto 0\}}
		}
	\\
	\tra{\silent}
	& \bufcore{[(x,1), (y,2), (z,3)]}{
		\cmdregs{\Skip}
		          {\{r \mapsto 2\}}
		}
\end{array}
\end{equation*}
\end{minipage}
\end{wrapfigure}
Consider a configuration comprised of 
three \Writes followed by a single \Read,
with
all variables initially zero. 
One possible execution, shown to the right,
is for all three \Writes to be placed in the write buffer via \refoprule{x86-write-buf}, leaving only the \Read of $y$ yet to execute.
As a \Write of $y$ can be found in the write buffer, the \Read $(r\asgn y)$ takes the value from the write buffer rather than from 
shared memory by \refoprule{x86-read-buf}.
Hence, the process reads $y = 2$, though the value of $y$ in shared memory may be different.

\end{multiexmp}

In \refoprule{x86-write-buf}, when a core
issues a write via rule \refoprule{store}, rather than interacting directly with the shared memory it is appended to the end of the write buffer. 
Because this has no effect on the shared memory, it is a silent transition with no effect on the rest of the system.

In \refoprule{x86-write-mem}, the oldest \Write-event in a write buffer can be ``flushed'' to the shared memory. 
This transition is labelled by the \Write-event that was earlier ``delayed'' by Rule \refoprule{x86-write-buf}.

In \refoprule{x86-read-mem}, a core attempts to read the shared variable $x$, where its write buffer contains no pending \Write to $x$. The core reads directly from the shared memory, making a transition labelled by a \Read-event.  
Note that this \Read interacts with the shared memory earlier than any pending \Writes in the write buffer, despite being program-ordered-after the corresponding stores.

In \refoprule{x86-read-buf}, a core reading the shared variable $x$ finds a corresponding pending \Write to $x$ in its write buffer. 
Then the core locally takes the value from this (right-most) pending \Write. 
This ``bypasses'' the shared memory, and is a silent step with no interaction with the rest of the system.

In \refoprule{x86-$\silent$}, a core can change from one local configuration to another via a silent transition, without modifying the write buffer. This occurs when
the core itself makes a silent transition, for instance, via rule \refoprule{skip}.

In \refoprule{x86-fence}, a $\ffence$ instruction can proceed, provided the write buffer is empty. 
This essentially forces all pending \Writes to be flushed (via \refoprule{x86-write-mem}) to the shared memory, before further instructions in $c$ can be issued.
A fence therefore ensures that \Reads after the fence occur strictly after the \Writes before the fence.




Combining the write-buffer rules (\reffig{rules-buffers}) with the \SeqCst rules (\reffig{rules}), we obtain an operational
description of TSO-style memory models, including that of x86 architectures.

The built-in ``delay'' induced by the write buffers result in the weak behaviour observed in the store-buffer litmus test (\refexmp{operat-tso-1}).
A standard view of weaker memory models is to have a single write buffer per variable, allowing independent writes to reorder within a single core,
and to extend with a read buffer per variable, allowing reordering of reads; in essence, the pipeline itself serves as all of these facilities at once.
The same set of rules can also be used to demonstrate forwarding; see \refexmp{operat-bypassing}.

In this presentation we have given each core its own write buffer;
an alternative approach 
is to consider the write buffers to be part of a ``storage subsystem'', separate to the cores;
in that structure each core effectively operates sequentially, irrespective of the memory model.
The memory model itself is wholly defined by the storage subsystem, which could include further buffers and cache mechanisms.


\subsection{Operational Descriptions of More Complex Models}
\labelsect{mechanistic-further}

The write buffer concept neatly captures the essence of the TSO (x86) memory model -- writes can be delayed in a buffer, keeping their
relative order but allowing reads to proceed early.  One way to extend the concept to related but weaker memory models, such as PSO (see \refsect{prelim-hard}),
is to use one write buffer per variable.  Then, writes to a variable $x$ are kept in order, but not necessarily with writes to another variable $y$.
In this setting, a fence must wait for all write buffers to be empty.  Extending the concept further to individual
load buffers (separate to the
write buffers) may work for the weaker-again models of Arm and RISC-V.  

Below, we outline how some of these more
complex aspects of weak memory may be approached through an explicit model of a hardware pipeline as described in
\refsect{pipelines}.
However, a focus on microarchitectural details
may become cumbersome to
work with formally.
For this reason many researchers focus on formal representations that abstract away from microarchitectural mechanisms, and focus instead on key
relationships between operations across multiple processes in the system.  
This is
typically referred to as an \emph{axiomatic} approach,
to which we turn our attention in \refsect{declar}.
Operationally one of the best-known approaches 
is the 
Promising Semantics \cite{kang-17,promising-2.0,promising-arm}, as discussed in \refsect{others-timestamp}, 
which combines an operational semantics with global axioms.

\ourparagraph{Arm and RISC-V}

\newcommand{\pline}{\mathtt{p}}
\newcommand{\loadop}[1]{\mathsf{load}(#1)}

The TSO memory model keeps the ordering amongst all stores, to any variables.  A key relaxation of this is to allow stores to different variables within a process to execute out of order.  Operationally, this can be
instituted by allowing stores to be propagated to main memory even when not at the head of the buffer, as 
given by the following rule for Arm, which is similar to \refoprule{x86-write-mem}.
\[
	\begin{array}{l}
  \text{(Arm-write-mem)} 
  	\quad
  	\Rule{
  		\lnot\exists (x,\_) \in \pline_1
		\quad
  		\lnot\exists \loadop{x,\_} \in \pline_1
		\quad
  		\lnot\exists \ffence \in \pline_1
	}{
		\bufcore{\pline_1 \cat (x,v) \cat \pline_2}{\lcfg} 
	   	\ttra{\WriteEvent{}{x}{v}} 
	   	\bufcore{\pline_1 \cat \pline_2}{\lcfg}
	}
	\end{array}
\]

In this case we assume \emph{all} instructions, not just writes, are fetched into the pipeline (generalising the
buffer), similarly to \refoprule{x86-write-buf}.
Provided that there are no writes to $x$ (of the form $(x,\_)$), reads of $x$ (of the form $\loadop{x,\_}$), 
or fences in the section of the pipeline preceding the write in question ($\pline_1$),
then the write can be issued.
A similar rule can be constructed for reads.  These constraints encode those of
\refdefn{order-deps}, modified by fences as in \refexmp{fence-tags},
and allow the weaker behaviours shown in \refexmp{sb-pipeline}. 
A more complete operational description of pipelines that conform
to the Arm memory model, and includes branches and jumps, is given in \cite{pipelines-cav-25}.

\ourparagraph{Release/Acquire}
The release/acquire concepts \cite{ReleaseConsistency90}, as introduced in 
\refeqn{first-example} and \refexmp{fence-tags-2},
can also be described operationally.
Continuing the generalisation of write buffers to pipelines, an operation with
release semantics can be executed only when at the head of the pipeline; i.e., when all other prior operations (w.r.t\ the program order) have completed. 
A load with acquire semantics prevents later memory operations from completing earlier; hence we extend the constraints of \refoprule{Arm-write-mem} to capture this possibility.
\[
  \text{(release-write)} ~~
  	\Rule{
	}{
		\bufcore{\relmod{(x,v)} \cat \pline}{\lcfg} 
	   	\ttra{\WriteEvent{}{x}{v}} 
	   	\bufcore{\pline}{\lcfg}
	}
\]
\[
  \text{(RA-write-mem)} ~~
  	\Rule{
  		\lnot\exists (x,\_) \in \pline_1
		\quad
  		\lnot\exists \loadop{x,\_} \in \pline_1
		\quad
  		\lnot\exists \ffence \in \pline_1
		\quad
		\boxed{
  		\lnot\exists \acqmod{\loadop{\_,\_}} \in \pline_1
		}
	}{
		\bufcore{\pline_1 \cat (x,v) \cat \pline_2}{\lcfg} 
	   	\ttra{\WriteEvent{}{x}{v}} 
	   	\bufcore{\pline_1 \cat \pline_2}{\lcfg}
	}
\]
Rule \refoprule{release-write} is similar to \refoprule{x86-write-mem} in that a release write
is issued only when at the head of the pipeline.
Rule \refoprule{RA-write-mem} extends \refoprule{Arm-write-mem} with an extra condition, which
we highlight with a box.
This additional constraint means that an earlier acquire load of \emph{any} variable prevents a later write from proceeding.
For the stronger form of release/acquire consistency in \cite{ReleaseConsistency90}
an acquire load additionally requires the absence of release stores in the earlier pipeline, 
which can be described syntactically similarly to the boxed constraint of \refoprule{RA-write-mem}.

\ourparagraph{Atomic operations}
Instruction types such as swap, compare-and-swap, fetch-and-increment, etc., combine read and write operations atomically.
These can be described using the concept of a \emph{global lock} which prevents a core from performing a
memory operation while another core holds the lock \cite{x86-TSO}. 
For instance, the atomic swap instruction, $\swap{\xx}{\xv}$, updates the shared variable $\xx$ to
hold the value in register $\regr$, and updates $\regr$ to the initial value of $\xx$; this instruction can be defined as follows.
\[
	\swap{\xx}{\regr} \equiv
		\lock \scomp \xtmp := \xx \scomp \xx := \regr \scomp \regr := \xtmp  \scomp \unlock
\]
Conceptually, the $\lock$ operation sets a flag that prevents other cores from carrying out any read or write, until $\unlock$ed.
The global $\lock$ can be specialised to per-variable locks, $\lockx$/$\unlockx$, to allow the reordering between accesses of independent locations,
as in Arm and RISC-V, while maintaining apparent atomicity per operation.  
The $\lockx$ operation may also be weakened to allow concurrent cores to read $\xx$ while it is locked.

%% file: declarative.tex
\newcommand{\predicate}[1]{\mathrm{#1}}
\newcommand{\acyclic}{\predicate{acyclic}}
\newcommand{\Empty}{\predicate{empty}}
\newcommand{\Irreflexive}{\predicate{irreflexive}}

\newcommand{\relcomp}{\mathop{;}}

\newcommand{\mydraw}{\draw [-Latex]}

\newcommand{\acquirefence}{\Rel{\textcolor{teal}{acqfence}}}
\newcommand{\loadfence}{\Rel{\textcolor{teal}{loadfence}}}

\section{Introduction to Axiomatic Formalisms}
\labelsect{declar}

The semantics of the previous section, based on write buffers, incorporates the definition of the memory model
into the rules governing of the structures of the system.  One of the more common approaches to formalising weak memory
behaviours is to abstract from mechanisms and instead describe properties of traces of the system.
In this section we provide an introduction to this common and effective style via a simplified version of the x86 
(Total Store Order) memory model.  
The extension of these concepts to more complex models is outlined in
\refsect{declar-further}; see also \refsect{declar-survey}.

\paragraph{Fundamental notation and concepts}

An axiomatic formalism specifies a memory model in terms of axioms that forbid certain executions of a concurrent program.
The concurrent program is represented as an \emph{execution graph} (\refdefn{exec-graph}), which comprises the set $E$ of \emph{memory events}, and a set of binary relations on $E$.
A memory event comes in the following types.
\begin{itemize}
  \item $\InitEvent{x}{v}$ denotes an initialisation-event that initialises the location $x$ to the value $v$.
  \item $\ReadEvent{i}{x}{v}$ denotes a read-event issued by Thread $i$ and reads the value $v$ from the location $x$.
  \item $\WriteEvent{i}{x}{v}$ denotes a write-event issued by Thread $i$ and writes the value $v$ to the location $x$.
  \item A \Fence event corresponds to a fence-instruction.
\end{itemize}
The binary relations are the subject of \refsect{declar-basic}. Some of these relations are common to all memory models; some are specific to individual memory models; some are specific to instruction-types on particular hardware-platforms;
some are derived from other relations using relational operators such as
\emph{converse} $(R^{-1})$,
\emph{composition} $(R_1 ; R_2)$,
\emph{intersection} $(R_1 \cap R_2)$,
\emph{union} $(R_1 \cup R_2)$, and
\emph{transitive closure} $(R^{+})$,
among others.

Axiomatic formalisms also make use of various restrictions of the identity-relation. Given a set $S$ of memory events, $[S]$ denotes $\{(e,e) \mid$ for every event $e \in S$ in the execution graph$\}$. 

Given an execution graph, an axiom on some relation $R$ takes one of the following forms, where the initialisation-events are excluded.
\begin{itemize}
  \item $\acyclic(R)$:
  The memory events can be arranged into a linear ordering that preserves $R$.

  Such a linear ordering is a ``topological ordering'' of the directed graph of $R$, 
  
  and it is well known that there exists a topological ordering if and only if the directed graph contains no directed cycle.

  \item $\Empty(R)$:
  No two memory events can be related by $R$.

  \item $\Irreflexive(R)$:
  No memory event can be related by $R$ to itself.
\end{itemize}

\subsection{Fundamental Relations}
\labelsect{declar-basic}

\reftable{declar-defs} summarises the key relations that will be covered in the rest of this section. Some of the relations are derived, as
indicated in \reffig{digraph-of-relations}.
In what follows, we use the variable convention where
$w$ denotes a write- or initialisation-event,
$r$ denotes a read-event,
and $e$ denotes any type of memory event.


\begin{table}[t]
\caption{Key relations on memory events}
\labeltable{declar-defs}
\vspace{-4mm}
\begin{tabular}{|lll|lll|}\hline
  & standard & & & alternatives & \\\hline\hline
    {\sf loc} & same-location & Def.~\ref{def:helper-relations} &
    & & \\\hline
    {\sf ext} & external & Def.~\ref{def:helper-relations} &
    & & \\\hline
    \po & program order & Def.~\ref{def:program-order} &
    {\sf sb} & sequenced-before & \citep{batty-11} \\\hline
    \poloc & \po per location & Def.~\ref{def:poloc} &
    & & \\
    & $\po \cap {\sf loc}$ & & & & \\\hline
    \fence & fence relation & Def.~\ref{def:fence-relation} &
    & & \\
      & $\po ; [F] ; \po$ & & & & \\\hline
    \co & coherence order & Def.~\ref{def:coherence-order} &
    {\sf mo} & modification order & \citep{batty-11} \\
    & & &
    {\sf ws} & write serialization & \citep{alglave-12} \\\hline
    \rf & reads-from & Def.~\ref{def:writes-to} &
    {\sf wt} & writes-to & \\\hline
    \fr & from-read & Def.~\ref{def:reads-before} &
    {\sf rb} & reads-before & \\
        & $\wt^{-1} ; \mo$ & & & & \\\hline\hline
    \com & communication & Def.~\ref{def:com-ca} & & & \\
         & $\mo \cup \wt \cup \rb$ & & & & \\\hline
    \ca & coherence-after & & & & \\
        & $\mo \cup \rb$ & & & & \\\hline
    {\sf eco} & extended coherence order & & & & \\
        & $\com^{+}$ & & & & \\\hline
    {\sf hb} & happens-before &  &
    & & \\\hline
    {\sf sw} & synchronises-with &  &
    & & \\\hline\hline
    \wte & external reads-from & Def.~\ref{def:wte} &
    {\sf wte} & external writes-to & \\
        & $\rf \cap {\sf ext}$ & & & & \\\hline
\end{tabular}
\end{table}


\begin{figure}[t]
  \centering
  \begin{tikzpicture}[
      every edge/.style = {draw, ->},
      every edge quotes/.style = {} ]
    \node (rf)  at ( 0, 2) [draw,rectangle] {\wt};
    \node (co)  at ( 2, 2) [draw,rectangle] {\mo};
    \node (fr)  at ( 1, 1) [draw,rectangle] {\rb};
    \node (fre) at ( 1, 0) [draw,rectangle] {\fre};
    \node (com) at ( 0, 0) [draw,rectangle] {\com};
    \node (ca)  at ( 2, 0) [draw,rectangle] {\ca};
    \node (coe) at ( 3, 1) [draw,rectangle] {\coe};
    \mydraw (rf) to (fr);
    \mydraw (co) to (fr);
    \mydraw (co) to (coe);
    \mydraw (co) to (ca);
    \mydraw (fr) to (ca);
    \mydraw (rf) to (com);
    \mydraw (fr) to (com);
    \mydraw [bend right] (co) to (com);
    \mydraw (fr) to (fre);
    \node (po)    at (-3, 2) [draw,rectangle] {\po};
    \node (loc)   at (-1, 2) [draw,rectangle] {\sf loc};
    \node (poloc) at (-2, 1) [draw,rectangle] {\poloc};
    \mydraw (po)  to (poloc);
    \mydraw (loc) to (poloc);
  \end{tikzpicture}
  \caption{Common memory-event relations and their relationships. An arrow from Node $A$ to Node $B$ means that Relation $A$ is used to define Relation $B$.}
  \labelfig{digraph-of-relations}
  \Description{TODO}
\end{figure}


\begin{definition}[loc and ext]
\labeldefn{helper-relations}
The \emph{same-location} relation {\sf loc} consists of pairs $(e_1, e_2)$ where $e_1$ and $e_2$ access (\ie read from or write to) the same memory location.

The \emph{external} relation {\sf ext} consists of pairs $(e_1, e_2)$ where $e_1$ and $e_2$ are issued by different threads.
\end{definition}

\begin{definition}
\labeldefn{program-order}
The \emph{program order} \po is a disjoint union of total orders on all events excluding initialisation-events.

It is specified by the program text, with the requirement that $(e_1,e_2) \in \po$ if and only if the memory events $e_1$ and $e_2$ are issued by the same process.

The meaning of $(e_1,e_2) \in \po$ is that the instruction corresponding to $e_1$ precedes the instruction corresponding to $e_2$ in the program text.
\end{definition}

\begin{definition}
\labeldefn{poloc}
The \emph{program order per-location} $\poloc$ is defined as $\po \cap {\sf loc}$. That is, it consists of pairs $(e_1, e_2) \in \po$ where $e_1$ and $e_2$ access the same memory location.
\end{definition}

The preservation of $\poloc$ is the one of the criteria of the ``coherence'' property (\refdefn{coherence}), which corresponds to the dependencies in processor pipelines (\refdefn{order-deps}).

\begin{definition}
\labeldefn{fence-relation}
The \fence relation is then defined as $\po ; [F] ; \po$, where $F$ is the set of all \Fence events in an execution graph.

Recall that $[F]$ is the identity-relation restricted to the set $F$, hence
\fence consists of pairs $(e_1,e_2)$ where there exists a fence $f$ such that $(e_1,f)\in\po$ and $(f,e_2)\in\po$.
\end{definition}


\begin{definition}
\labeldefn{coherence-order}
The \emph{coherence order} \co is a disjoint union of total orders on initialisation- and write-events.

It is specified by the program execution (as opposed to \po, which is specified by the program text), with the requirement that $(e_1,e_2) \in \co$ if and only if (1) $e_1$ is an initialisation- or write-event, (2) $e_2$ is a write-event, and (3) $e_1$ and $e_2$ write to the same location.

The meaning of $(e_1,e_2) \in \co$ is that $e_1$ takes place globally before $e_2$.
\end{definition}

\begin{definition}
\labeldefn{writes-to}
\labeldefn{reads-from}
The relation \rf is specified by the program execution, and consists of pairs $(w,r)$ where $w$ (an initialisation- or write-event) and $r$ (a read-event) accesses the same location and have the same value. (That is, $r$ reads the value written by $w$.)

Note that the converse of a valid \rf must be functional. Indeed, $(r,w) \in \rf^{-1}$ means that $r$ reads from $w$, and every read-event $r$ must have a unique ``source''. This property underpins \refdefn{reads-before} later.
\end{definition}

On the name ``reads-from'', here we follow the recent work by Arm~\citep{alglave-21}, where a typical pair $(w,r) \in \rf$ means ``$r$ reads from $w$'', with $w$ occurring before $r$. Some may find it more intuitive to think of the relation \rf as \emph{writes-to}. Indeed, $(w,r) \in \rf$ means that $w$ \emph{writes to} the value read by $r$.

The relations \po, \co and \rf are the ``base relations''. 
The set of memory events and these base relations constitute an ``execution graph'', also known as a ``candidate execution''~\citep{alglave-21}.

\begin{definition}
\labeldefn{exec-graph}
An \emph{execution graph} comprises the set of memory events, and \po, \co and \rf.
\end{definition}


Next we present the key \emph{derived relations}, extracted from a given execution graph.

\begin{definition}
\labeldefn{reads-before}
The \emph{from-read} relation \fr is defined to be $(\wt^{-1} \relcomp \mo)$.
\end{definition}

A typical pair in \fr has the form $(r,w)$, where $r$ and $w$ access the same location, and there exists another \Write $w'$ such that $(w',w)\in\mo$ and $(w',r)\in\wt$.

One intuition behind \fr is the following. Whereas the reads-from relation \wt places a \Read $r$ strictly after the \Write from which $r$ gets its value, the from-read relation \fr places each \Read strictly before the subsequent \Write that changes the relevant location. It thus ``fills-in'' the order of operations on a single variable.
Some may find it more intuitive to think of \fr as \emph{reads-before}, since $(r,w) \in \rb$ essentially means that $r$ \emph{reads} a location right \emph{before} $w$ changes it.

For instance, let $w_1$ and $w_2$ be ordered \Writes to $x$,
and $r$ is a \Read of $x$ that takes its value from $w_1$. 
Then, the coherence order on $x$ is $\{(w_1,w_2)\}$,
and the reads-from relation is $\{(w_1,r)\}$.
Hence, the derived from-read relation is $\{(r,w_2)\}$. 
This makes explicit that $r$ must occur before $w_2$,
or else it would have read a different value.
Combining all these, the order of the events on $x$ is thus
$w_1, r, w_2$.


\paragraph{Other Common Relations}
Below are some relations that are common in the literature but are not used in the examples in the rest of this paper. We include them here for completeness.

\begin{definition}[com and ca]
\labeldefn{com-ca}
The relations \mo, \wt and \rb above are sometimes combined into the following abbreviations.
\begin{itemize}
  \item
  The \emph{communication order}
  $\com = \mo \cup \wt \cup \rb$. 
  \item 
  The \emph{coherence-after} relation
  $\ca = \mo \cup \rb$.
  \item
  The \emph{extended coherence order} ${\sf eco} = \com^{+}$ -- defined to be the transitive closure of \com~\cite{lahav-17}.
\end{itemize}
\end{definition}

In addition, software memory models make use of the following two relations to enforce ordering.
\begin{itemize}
  \item
  The \emph{synchronises-with} relation ({\sf sw}) -- induced by certain synchronisation constructs.
  \item
  The \emph{happens-before} relation ({\sf hb}) -- contains \po and {\sf sw}.
\end{itemize}
These are induced by specific language features, such as the ``release/acquire'' tags, and ``volatile'' variables in C and Java.

\subsection{Axiomatic Definition of \SeqCst}
\labelsect{declar-sc}

\begin{definition}[Coherence]
\labeldefn{coherence}
An execution graph yields a \emph{coherent} execution if:
\[ \acyclic(\poloc \cup \co \cup \rf \cup \fr) .\]

In other words, a possible execution is \emph{coherent} if it preserves this relation.
\end{definition}

A memory model is said to satisfy the \emph{coherence} property if all possible executions of any execution graph is coherent. The coherence property is also known as the \emph{uniproc} property~\citep{alglave-12}, \emph{SC per location}~\citep{alglave-14}, \emph{SC per variable}~\citep{alglave-cousot-17} or \emph{internal visibility}~\citep{alglave-21}.

The coherence property is satisfied by most memory models. 
It ensures that possible executions do not contain ``invalid'' pairs of events such as $\WriteEvent{A}{x}{2}$ and then $\ReadEvent{B}{x}{3}$. 
Indeed, once Thread $A$ has written 2 to $x$ globally, Thread $B$ should not be able to read 3 from $x$.
(See also: the discussion following \refdefn{tso-axiom}.)

\begin{definition}[SC Axiom]
\labeldefn{sc-axiom}
An execution graph yields a \emph{sequentially consistent} execution if:
\[ \acyclic(\po \cup \co \cup \rf \cup \fr) .\]
\end{definition}

Note that the coherence property follows from this definition because $\poloc \subseteq \po$.
Examples~\ref{exmp:declar-sc} and~\ref{exmp:declar-sc-2} now demonstrate how this definition is applied.

\OMIT{ 
\begin{table}[t]
  \caption{Simple strong and weak memory model specifications}
  \labelfig{mm-specs}
  \vspace{-2mm}
  \begin{tabularx}{\textwidth}{X}
  \hline

  \begin{definition}[Sequentially Consistent Memory Model]
  \labeldefn{sc-axioms}\labeldefn{seq-cst}
  Any execution candidate of program $c$ that is \emph{acyclic} and is consistent with the program order ($\po$) of $c$ 
represents a sequentially consistent execution.
  \end{definition}

  \begin{definition}[Weak Memory Model]
  \labeldefn{weak-axioms}
  Any execution candidate of program $c$ that is \emph{acyclic} and is consistent with program order per-location ($\poloc$) and respects
the fences ($\fence$) of $c$ represents a minimally weak execution.
  \end{definition}

  \\\hline
  \end{tabularx}
\end{table}

A sequentially consistent execution (\refdefn{sc-axioms}) observes all events in the system to satisfy the original program order. The modification order of variables (\mo) connects writes from thread to thread in a globally consistent, acyclic order, and the writes-to relation (\wt) connects reads to the corresponding write. Crucially, the resulting relation must be \emph{acyclic}, otherwise this would represent an execution with no valid traces.

We may weaken sequentially consistent executions so that rather than respecting program order in total they respect some minimal relations, in particular, operations on each variable are kept in order (\poloc) and fences are respected (\fence).  These two relations are subsets of the original program order, and as described above are required to maintain sequential semantics when considered in a single threaded context, and to allow a program to reinstate order if necessary. Essentially ARM and RISC-V require only these constraints (for our simplified language), while x86 lies between the two; we investigate it in more detail in the next section.
} 

\begin{multiexmp}

\exmpheading{Reachable state of \SB under SC}
\labelexmp{declar-sc}
Consider the \SB litmus test (\refexmp{store-buffer}) with the postcondition $r_1 = 1 \land r_2 = 0$. This corresponds to the execution graph on the left-hand side below, where the solid unlabelled edges are the pairs in \rf. The dashed unlabelled edge $(b_2,a_1)$ is the sole edge in \fr, which is the composition of the reversed \rf-edge $(b_2,I_x)$ and the \co-edge $(I_x,a_1)$.

\quad
This digraph now consists of relations relevant to SC (Def.~\ref{def:sc-axiom}), and contains no cycle, so we conclude that this execution graph yields a sequentially consistent execution. Indeed, $\seqT{b_1 \ssep b_2 \ssep a_1 \ssep a_2}$ is one such execution (omitting the initialisation-events).

\begin{center}
  \begin{tikzpicture}[
      every node/.style = {font=\scriptsize},
      every edge/.style = {draw, ->},
      every edge quotes/.style = {auto, font=\scriptsize}]
    \node (0x) at (-1.2, 2.4) [draw] {$I_x = \InitEvent{x}{0}$};
    \node (0y) at ( 1.2, 2.4) [draw] {$I_y = \InitEvent{y}{0}$};
    \node (a1) at (-1.2, 0.8) [draw] {$a_1 = \WriteEvent{a}{x}{1}$};
    \node (a2) at (-1.2,-0.8) [draw] {$a_2 = \ReadEvent{a}{y}{1}$};
    \node (b1) at ( 1.2, 0.8) [draw] {$b_1 = \WriteEvent{b}{y}{1}$};
    \node (b2) at ( 1.2,-0.8) [draw] {$b_2 = \ReadEvent{b}{x}{0}$};
    \draw (0x) edge["\co"] (a1);
    \draw (0y) edge["\co"] (b1);
    \draw (a1) edge["\po"] (a2);
    \draw (b1) edge["\po"] (b2);
    \draw (0x) edge[red] (b2);
    \draw (b1) edge[red] (a2);
    \draw (b2) edge[orange,dashed] (a1);
  \end{tikzpicture}
  \qquad
  \begin{tikzpicture}[
      every node/.style = {font=\scriptsize},
      every edge/.style = {draw, ->},
      every edge quotes/.style = {auto, font=\scriptsize}]
    \node (0x) at (-1.2, 2.4) [draw] {$I_x = \InitEvent{x}{0}$};
    \node (0y) at ( 1.2, 2.4) [draw] {$I_y = \InitEvent{y}{0}$};
    \node (a1) at (-1.2, 0.8) [draw] {$a_1 = \WriteEvent{a}{x}{1}$};
    \node (a2) at (-1.2,-0.8) [draw] {$a_2 = \ReadEvent{a}{y}{0}$};
    \node (b1) at ( 1.2, 0.8) [draw] {$b_1 = \WriteEvent{b}{y}{1}$};
    \node (b2) at ( 1.2,-0.8) [draw] {$b_2 = \ReadEvent{b}{x}{0}$};
    \draw (0x) edge["\co"] (a1);
    \draw (0y) edge["\co"] (b1);
    \draw (a1) edge["\po"] (a2);
    \draw (b1) edge["\po"] (b2);
    \draw (0y) edge[red] (a2);
    \draw (0x) edge[red] (b2);
    \draw (b2) edge[orange,dashed] (a1);
    \draw (a2) edge[orange,dashed] (b1);
  \end{tikzpicture}
\end{center}

\exmpsep

\exmpheading{Unreachable state of \SB under SC}
\labelexmp{declar-sc-2}

Now consider the same litmus test, but instead with the postcondition $r_1 = r_2 = 0$. This corresponds to the execution graph on the right-hand side above. Again, the solid unlabelled edges are the pairs in \rf, and the dashed unlabelled edges are the pairs in \rb. This time, there is a cycle $a_2 \rightarrow b_1 \rightarrow b_2 \rightarrow a_1 \rightarrow a_2$,
so according to \refdefn{sc-axiom}, 
any final state with $r_1 = r_2 = 0$ is not reachable under SC. 

\end{multiexmp}

\subsection{Axiomatic Definition of TSO}\labelsect{declar-tso}

Compared to \SeqCst, the Total Store Order memory model allows two additional phenomena: (1) \Reads can take effect earlier than \Writes, making \Write-\Read pairs appear reordered; and (2) a process can read directly from its own \Writes before they are observed globally. This latter phenomenon is known as \emph{forwarding} (see \refsect{forwarding}).

As instructions can appear reordered due to (1) above, fence instructions (see \refsect{fences}) are used to prevent any undesirable reordering.
In this section, we introduce the axiomatic concepts for reasoning about \Write-\Read reordering, \Fence instructions and forwarding.

To capture the phenomenon of \Write-\Read reordering, we define the following four relations to filter certain types of pairs from the program order.

\begin{definition}[Local Event Relations]
\labeldefn{rr-rw-wr-ww}
\begin{align*}
  \RR &= \{ (r, r) \in \po \} &
  \RW &= \{ (r, w) \in \po \} &
  \WW &= \{ (w, w) \in \po \} &
  \WR &= \{ (w, r) \in \po \}
\end{align*}
%
%
\end{definition}

Given a pair $(w,r)\in\wt$ with $w$ and $r$ issued by the same thread,
forwarding (\refsect{forwarding}) can be phrased as: 
$r$ reads the value written by $w$, but appears to take effect earlier than $w$ in the global order (violating the principle of the reads-from relation).
Hence, forwarding is described axiomatically by relaxing \wt, and dropping those \Write-\Read pairs in \wt whose events are issued by the same thread.
This is described by the ``external'' reads-from relation.

\begin{definition}[rfe]\label{def:wte}
The \emph{external writes-to} relation, written \wte, consists of the pairs $(w,r)\in\wt$ where $w$ and $r$ are issued by \emph{different} threads.
\end{definition}


\begin{definition}[PPO of TSO]
The \emph{preserved program order of TSO} is $\ppotso = \RR \cup \RW \cup \WW$.
\end{definition}

\begin{definition}[TSO Axioms]\labeldefn{tso-axiom}
An execution graph yields a possible execution under TSO if:
\[ \acyclic(\poloc \cup \co \cup \wt \cup \rb)
   \text{\quad and \quad}
   \acyclic(\ppotso \cup \fence \cup \co \cup \wte \cup \rb) .\]
\end{definition}

The first axiom is the coherence axiom (\refdefn{coherence}), 
and the second is the PPO axiom.
TSO explicitly requires the coherence axiom,
as illustrated by \refexmp{tso-coherence}.
Note that this is in contrast to SC (\refdefn{sc-axiom}) 
where the coherence axiom is implied.

Examples~\ref{exmp:declar-sb-reachable-tso},~\ref{exmp:declar-sb-tso-fence} and~\ref{exmp:declar-forwarding}
show how the TSO axioms are applied in practice.


\begin{exmp}[Coherence Axiom of TSO]
\labelexmp{tso-coherence}

To see why TSO (\refdefn{tso-axiom}) requires both the coherence axiom and the PPO axiom,
consider the one-thread execution graph
corresponding to $(x \asgn 1 \scomp r \asgn x)$. 

\begin{center}
\begin{tikzpicture}[
  every node/.style = {font=\scriptsize},
  every edge/.style = {draw, ->},
  every edge quotes/.style = {font=\scriptsize}]
  \node (0) at (-3, 0) [draw] {$\InitEvent{x}{0}$};
  \node (1) at ( 0, 0) [draw] {$\WriteEvent{}{x}{1}$};
  \node (2) at ( 3, 0) [draw] {$\ReadEvent{}{x}{0}$};
  \draw (0) edge[above,"\co"] (1);
  \draw (1) edge[above,"\poloc"] (2);
  \draw (0) edge[above,"\rf",bend left=20] (2);
  \draw (2) edge[below,"\fr",bend left=20] (1);
\end{tikzpicture}
\end{center}

Without the coherence axiom in \refdefn{tso-axiom}, 
the PPO axiom concerns only \co and \fr. 
The corresponding digraph is acyclic, 
meaning that there exists a possible execution, and in every execution, $\ReadEvent{}{x}{0}$ must precede $\WriteEvent{}{x}{1}]$. 
This, however, is not coherent, as it alters the fundamental semantics of the original program.
To rule out such incoherent executions, the coherence axiom takes the \poloc edges into account, leading to the digraph $(\poloc \cup \co \cup \wt \cup \rb)$, which indeed has a cycle, and ensures that this nonsensical execution graph yields no possible execution.

\end{exmp}

\begin{multiexmp}

\exmpheading{Reachable state of \SB under TSO}
\labelexmp{declar-sb-reachable-tso}

Consider the SB litmus test with the postcondition $r_1 = r_2 = 0$. The corresponding execution graph is the same as that of \refexmp{declar-sc-2}, with $\fence = \emptyset$ and $\wte = \wt$.
Now as both $(\poloc \cup \mo \cup \wte \cup \rb)$ and $(\ppotso \cup \fence \cup \mo \cup \wte \cup \rb)$ are acyclic, this execution graph does admit an execution under TSO, namely $\seqT{a_2 \ssep b_2 \ssep b_1 \ssep a_1}$.


\begin{wrapfigure}{r}{0.42\textwidth}
  \begin{tikzpicture}[
      every node/.style = {font=\scriptsize},
      every edge/.style = {draw, ->},
      every edge quotes/.style = {font=\scriptsize}]
    \node (0x) at (-1.3, 2) [draw] {$I_x = \InitEvent{x}{0}$};
    \node (0y) at ( 1.3, 2) [draw] {$I_y = \InitEvent{y}{0}$};
    \node (a1) at (-1.3, 1) [draw] {$a_1 = \WriteEvent{a}{x}{1}$};
    \node (af) at (-1.3, 0) [draw] {\Fence};
    \node (a2) at (-1.3,-1) [draw] {$a_2 = \ReadEvent{a}{y}{0}$};
    \node (b1) at ( 1.3, 1) [draw] {$b_1 = \WriteEvent{b}{y}{1}$};
    \node (bf) at ( 1.3, 0) [draw] {\Fence};
    \node (b2) at ( 1.3,-1) [draw] {$b_2 = \ReadEvent{b}{x}{0}$};
    \draw (0x) edge[left,"\co"] (a1);
    \draw (0y) edge[left,"\co"] (b1);
    \draw (a1) edge[left,"\po"] (af);
    \draw (af) edge[left,"\po"] (a2);
    \draw (b1) edge[right,"\po"] (bf);
    \draw (bf) edge[right,"\po"] (b2);
    \draw (0x) edge[right,"\rf",bend left =10] (b2);
    \draw (0y) edge[left, "\rf",bend right=10] (a2);
    \draw (a1) edge[left, "\fence",bend right=60] (a2);
    \draw (b1) edge[right,"\fence",bend left =60] (b2);
    \draw (a2) edge[orange,right,"\rb"] (b1);
    \draw (b2) edge[orange,left, "\rb"] (a1);
  \end{tikzpicture}
\end{wrapfigure}

\exmpsep 

\quad

\exmpheading{Fences in \SB under TSO}
\labelexmp{declar-sb-tso-fence}

Now we show how fences prevent the \Write-\Read reordering of TSO. Inserting fences into the SB litmus test,
as in \refexmp{fence-tags}, results in the execution graph on the right.

\quad
As the relations \fence and \rb create a cycle, we can conclude that $(\ppotso \cup \fence \cup \co \cup \wte \cup \rb)$ is not acyclic, and hence no execution is possible.

\exmpsep 

\exmpheading{Forwarding under TSO}
\labelexmp{declar-forwarding}

Consider the program on the left below,\\ which generates the memory-events on the right. Here, the value written in $a_1$ is forwarded to $a_2$, but is not visible to $b_3$.

\begin{center}
  \begin{tabular}{c}
  $\{ x = 0 \land y = 0 \}$ \\
  \begin{tabular}{r||r}
    $x      := 1$ & $y      := 1$ \\
    $r_{ax} := x$ & $r_{by} := y$ \\
    $r_{ay} := y$ & $r_{bx} := x$
  \end{tabular} \\
  $\{ r_{ax} = 1 \land r_{ay} = 0 \land r_{by} = 1 \land r_{bx} = 0 \}$
\end{tabular}
\qquad
\begin{tabular}{ll}
  $I_x = \WriteEvent{ }{x}{0}$ \quad&\quad $I_y = \WriteEvent{ }{y}{0}$ \\
  $a_1 = \WriteEvent{a}{x}{1}$ \quad&\quad $b_1 = \WriteEvent{b}{y}{1}$ \\
  $a_2 =  \ReadEvent{a}{x}{1}$ \quad&\quad $b_2 =  \ReadEvent{b}{y}{1}$ \\
  $a_3 =  \ReadEvent{a}{y}{0}$ \quad&\quad $b_3 =  \ReadEvent{b}{x}{0}$
\end{tabular}
\end{center}

The corresponding execution graph is depicted separately below: one digraph contains the relations for the coherence property, and the other the relations for the PPO of TSO.

\begin{center}
\begin{tabular}{cc}
\begin{tikzpicture}[
  every node/.style = {font=\scriptsize},
  every edge/.style = {draw, ->},
  every edge quotes/.style = {font=\scriptsize}]
  \node (0x) at (-1, 2) [draw] {$I_x$};
  \node (0y) at ( 1, 2) [draw] {$I_y$};
  \node (a1) at (-1, 1) [draw] {$a_1$};
  \node (a2) at (-1, 0) [draw] {$a_2$};
  \node (a3) at (-1,-1) [draw] {$a_3$};
  \node (b1) at ( 1, 1) [draw] {$b_1$};
  \node (b2) at ( 1, 0) [draw] {$b_3$};
  \node (b3) at ( 1,-1) [draw] {$b_3$};
  \draw (0x) edge[left,"\mo"] (a1);
  \draw (0y) edge[right,"\mo"] (b1);
  \draw (0x) edge[right,"\wt"] (b3);
  \draw (0y) edge[right] (a3);
  \draw (a1) edge[left,"\poloc~\wt"] (a2);
  \draw (b1) edge[right,"\poloc~\wt"] (b2);
  \draw (a3) edge[orange] (b1);
  \draw (b3) edge[orange,"\rb"] (a1);
\end{tikzpicture}
&
\begin{tikzpicture}[
  every node/.style = {font=\scriptsize},
  every edge/.style = {draw, ->},
  every edge quotes/.style = {font=\scriptsize}]
  \node (0x) at (-1, 2) [draw] {$I_x$};
  \node (0y) at ( 1, 2) [draw] {$I_y$};
  \node (a1) at (-1, 1) [draw] {$a_1$};
  \node (a2) at (-1, 0) [draw] {$a_2$};
  \node (a3) at (-1,-1) [draw] {$a_3$};
  \node (b1) at ( 1, 1) [draw] {$b_1$};
  \node (b2) at ( 1, 0) [draw] {$b_3$};
  \node (b3) at ( 1,-1) [draw] {$b_3$};
  \draw (0x) edge[left,"\mo"] (a1);
  \draw (0y) edge[right,"\mo"] (b1);
  \draw (a2) edge[left, "\ppo"] (a3);
  \draw (b2) edge[right,"\ppo"] (b3);
  \draw (a3) edge[orange] (b1);
  \draw (b3) edge[orange,"\rb"] (a1);
\end{tikzpicture}
\end{tabular}
\end{center}

Both digraphs are acyclic, so by \refdefn{tso-axiom}, an execution is possible under TSO.

\end{multiexmp}

\subsection{Axiomatic Descriptions of More Complex Models}
\labelsect{declar-further}

For illustrative purposes, we have presented how axiomatic formalisms 
can define the straightforward SC memory model and a simplified version of TSO. 
In practice, axiomatic formalisms have been used to specify a wide range of behaviours related to 
more complex memory models and instruction types,
e.g., \cite{alglave-14,pulte-17,alglave-21}, as surveyed in \refsect{declar-survey}.
Below we give an indication of how these more complex problems are tackled axiomatically.

\newcommand{\addr}{\Rel{\textcolor{purple}{addr}}}
\newcommand{\data}{\Rel{\textcolor{purple}{data}}}
\newcommand{\ctrl}{\Rel{\textcolor{purple}{ctrl}}}

\ourparagraph{Arm and RISC-V}
Arm and RISC-V further weaken TSO by removing
the \RR, \RW, and \WW\xspace relations (\refdefn{rr-rw-wr-ww}) from the preserved program order (\refdefn{tso-axiom}).  
This weakening corresponds to the principles of pipelining in \refdefn{order-deps}.
This is straightforward to instrument and, because it is based on (subsets of) relations,
clearly shows a weakening, for which approaches based on microarchitectural mechanisms (\refsect{mechanistic-further}) would 
typically require more complex arguments.

Additional relations are required to handle aspects of assembler-level languages:
\data,
where the value assigned to a variable is dependent on previous instructions;
\addr, where the calculation of an accessed address in memory is dependent on previous instructions;
and \ctrl, to handle branch instructions, where, on Arm and RISC-V architectures for example,
later load instructions can be speculatively
executed before a branch condition is resolved, though later stores cannot.

Substantial parts of the Arm and RISC-V memory models concern specific instruction-types;
in particular, different types of fences and release-acquire constraints. 
For instance, RISC-V's \texttt{fence.r.r} instruction enforces order between loads, similar to Arm's 
load fence (\texttt{dmb.ld}).  

\ourparagraph{Release/Acquire} 
Let the set $\Rls$ include any operation with \emph{release} semantics (e.g., a release fence, or a store designated as a release).
Axiomatically, we require the preservation of the relation $\po ; \Rlsrel$,
so that any operation program-ordered before a release operation is guaranteed to have completed prior.
This allows a release operation to act as a ``flag'' that some update has occurred in main memory
(see, e.g., \refeqn{first-example} and \refexmp{fence-tags-2}).

Similarly, let the set $\Acq$ include any operation with \emph{acquire} semantics (e.g., an acquire fence, or a load designated as an acquire).
We then add $\Acqrel ; \po$, to the relations to preserve,
meaning that any operation program-ordered after an acquire operation will occur after.
In combination with a release constraint, this allows a form of message-passing communication via shared variables.

Finally, to maintain order between release and acquire locally, we require
$
	\Rlsrel ; \po ; \Acqrel
$.
(The Arm official model also includes requirements between release/acquire operations and RMWs.)

\OMIT{
If we let $\loadfence$ be the fence relation specific to load fences, similar to 
\fence (\refdefn{fence-relation}),
then the corresponding axiomatic constraint can be expressed simply as $[\tw{R}] \relcomp \loadfence \relcomp [\tw{R}]$.
The behaviour of release/acquire constraints can be similarly straightforwardly captured,
for instance, a \Clang-like \emph{acquire fence} that forces order
between loads and any later memory events
can be described by $[\tw{R}] \relcomp \acquirefence \relcomp [\tw{E}]$.
}

%

\ourparagraph{Atomic instructions}
Most hardware platforms support RMW instructions such as \emph{compare-and-swap} (CAS) and \emph{test-and-set} (TAS). 
Such an instruction reads from a memory location, performs some local computation involving the read value,
and finally writes to memory, all in one atomic step. 
When constructing an execution graph, an RMW instruction corresponds to both a \Read- and a \Write-event. The atomicity of such a pair of \Read and \Write can be enforced axiomatically,
i.e.,
the \rmw relation is a subset of \po, and relates a \Read and a \Write that are to occur atomically
according to some atomic instruction type.

Relations \fr and \co are used to filter \rmw events using {\sf ext} (\refdefn{helper-relations}) to obtain the pairs whose events are issued by different threads.
\[ \fre = \fr \cap {\sf ext} \qquad \coe = \co \cap {\sf ext} \]
Now, a pair $(r,w)$ belongs to the relation $(\fre \relcomp \coe)$ precisely when $r$ and $w$ are interleaved by a \Write issued by another thread.
In other words, $(r,w) \in (\fre \relcomp \coe)$ means that $r$ and $w$ are \emph{not} atomic.
Hence, to ensure that every pair in \rmw is atomic, we require that no pair in \rmw is in $(\fre \relcomp \coe)$, 
that is,
an execution graph enforces the RMW instructions if
$ \Empty(\rmw \cap (\fre \relcomp \coe)) $.
Further details and mechanised tool support can be found in the online repository of the Herd tool~\citep{herd7website,alglave-21}.


%% file: formalisations.tex

\section{Survey of Memory Model Formalisms}
\labelsect{formalisms}
\labelsect{formalisations}
\labelsect{representations}

In \refsects{operat}{declar} we gave detailed presentations for simplified versions of operational and axiomatic formalisms. In this section, we expand on the relevant literature, and also cover other approaches to representing memory models.

%

\subsection{Mechanistic Formalisms}
\labelsect{mechanistic-survey}

Mechanistic formalisms describe the effects of weak memory models with structures that operate in the manner of hardware mechanisms.
The best-known of such mechanisms is the \emph{local write buffers}
elucidated in \refsect{operat}.
Local write buffers describe the behaviour of TSO~\cite{sindhu-91-tso}, on which the x86 memory model is based~\cite{owens-09,sarkar-09,sewell-10}. 
As local write buffers are relatively straightforward to encode, they often serve as the foundation for verification under TSO. Namely, when each write buffer is modelled as a separate process, one can analyse the entire system under \SeqCst using existing tools~\cite{burk-08,travkin-13-lineariz,travkin-14-lineariz,wehrheim-travkin-15-haifa,travkin-wehrheim-16,linden-10,linden-11,linden-13,wmms-as-llvm-trans-16}.

The mechanistic formalism of \citet{boudol-09} is also based on write buffers that are not local but potentially hierarchical, making this formalism capable of expressing behaviours weaker than those of TSO.
Another recent variant of write buffers is \emph{read buffers}, proposed by \citet{abd-18-load-buffer-TSO} to improve various aspects of verification under TSO. 
Meanwhile, two other major hardware platforms -- Power~\cite{sarkar-11} and Arm~\cite{ModellingARMv8,pulte-17} -- also have their own mechanistic formulations, albeit much more complex than x86-TSO and harder to encode for verification tools.
A semantics based directly on pipelined and speculative execution, as outlined in \refsect{pipelines}, has been developed for Arm, RISC-V and x86
processors \cite{pipelines-cav-25}.

The main advantages of mechanistic formalisms is their \emph{intuition} and \emph{compositionality}: as they align closely with the actual mechanisms on processors, they give a local understanding of how weak memory behaviours arise, without needing to understand the system as a whole.
On the other hand, the definition of the memory model itself is integrated in the operational rules, rather than being a stand-alone definition.
It is also harder to show which behaviours \emph{cannot} happen without a full analysis of all traces, in comparison with axiomatic
approaches.
Finally, a mechanistic description does not necessarily coincide with the behaviour of a compiler.

\subsection{General Operational Formalisms: Timestamped Messages, Promises and Views}
\labelsect{others-timestamp}

Hardware-like mechanisms are insufficient for describing the behaviours of compilers -- the source of complexity for the C memory model especially. In order to reason about the C memory model operationally, many formulations make use of the notions of \emph{timestamped messages} and \emph{views}. 
Under weak memory models, a \Write does not become visible to other threads immediately. Thus, each \Write can be seen as a message that is broadcast by the issuing process, and it is up to the memory model that dictates \emph{when} this message becomes visible to \emph{which} other processes. 
Each process has a view of a subset of all the messages that have been sent out, and each message has a timestamp attached. By looking at the timestamps, a process updates its view according to the memory model.
For instance, instead of just $\Wev{x}{v}$, a \Write-event is now a pair $(\Wev{x}{v}, t)$, where $t$ is the time at which the event took place. The timestamp $t$ can be some future time, meaning that any subsequent \Read of $x$ with return-value $v$ must occur at some later time $t'$, \ie $t \leq t'$.
An early use of timestamps is provided in \cite{podkopaev-16};
we look at some other prominent formalisations in this vein below.

\paragraph{The ``Promising Semantics''}
In this framework, introduced in
\citet{kang-17}, each timestamped message is a \emph{promise} to fulfil a later \Write. This semantics was later simplified by \citet{promising-2.0}, and adapted to Arm and RISC-V by \citet{promising-arm}.
The semantics has been used as the basis for verification of non-trivial systems \cite{tao-21}.

A characteristic of the Promising Semantics is that
it is a \emph{multi-execution} formalism~\citep{mois-22},
as opposed to a \emph{single-execution} formalism.
In a single-execution formalism (e.g., that of \refsect{operat}),
one behaviour, e.g., a trace of memory operations, corresponds to one execution of the code. 
In the Promising Semantics a single behaviour comprises more information that just the sequence memory events,
and this extra information also covers multiple possible executions of the code.
The multi-execution characteristic can also be applied to partial-order-based representations
(see \refsect{others-pomsets}).

\paragraph{Views and Potentials}
Formalisms with timestamped messages and similar notions have since been developed on several fronts,
for example, \citet{doherty-19}, with subsequent works leading to program logics~\cite{dalvandi-20-og,wright-21-og,doherty-22-unify,dalvandi-22,dalvandi-22-og,bargmann-pso-23,dongol-23}.
\citet{lahav-boker-22} introduced the notion of \emph{potentials}. In contrast to timestamped messages which reason about \Stores that have occurred in the \emph{past}, potentials reason about possible \Stores in the \emph{future}.

Timestamped messages, views, and potentials enable general frameworks for handling a wide range of weak memory effects, including the lack of multicopy atomicity (\refsect{lmca}) and out-of-thin air behaviour (\refsect{prelim-soft}).
This flexibility is not without disadvantages, however: these formalisms are perhaps the most complex of those surveyed,
with the details of even small examples such as the store buffering litmus test being difficult to reason about \cite{mois-22}.

\subsection{Axiomatic Formalisms}
\labelsect{formalism-axiomatic}
\labelsect{declar-survey}

One of the most successful approaches to formalising memory models is via constraints on the allowed behaviours. This is referred to as the \emph{axiomatic} (or \emph{declarative}) approach in the literature \cite{alglave-14,sindhu-91-tso}.
Given a set of memory events (where each \Read has a return value), one starts with all possible permutations, and retains only those permutations that satisfy the \emph{axioms} of the particular memory model
(conceptually the opposite of first considering the set of behaviours allowed under \SeqCst and then expanding that set to capture the weak behaviours).

Due to its preciseness and relative conciseness, the axiomatic approach is widely adopted for the major memory models,
\eg x86-TSO~\cite{owens-09}, Arm~\citep{alglave-21}, RISC-V~\citep{RISC-V}, Java~\cite{manson-05}, and C~\citep{batty-11}.

Some of the most influential tools in this area are the domain-specific
language Cat and its associated litmus-test simulator Herd, introduced in \cite{alglave-14}.
Cat and Herd are utilised by Arm to formalise its memory model~\citep{alglave-21},
and Cat is also used in a number of model checkers for expressing other memory models (see \refsect{tools-check}).
An axiomatic formalisation is leveraged by the \emph{Check} suite of tools and techniques for investigating
correctness and security of microarchitectural implementations \cite{PipeCheck,PipeProof,CheckMate,CCICheck-2015,COATCheck-2016,TriCheck}.

The downside of axiomatic formalisms is that they are based on global reasoning about
completed executions, where every $\Read$ has a
value, and the relations $\mo$ and $\wt$ are supplied (see \refsect{declar}).  
As such, axiomatic formalisms tend not to be suited for reasoning about programs compositionally;
that is, one cannot form the set of behaviours of a system by considering individual processes in isolation.
The purely axiomatic approach has less applicability to theorem proving (however, see \cite{alglave-cousot-17,hammond-axsl-24}), so operational semantics (see \refsect{others-timestamp}) are sometimes developed in tandem, often with an underlying axiomatic flavour~\citep{doherty-22-unify,kang-17,promising-arm,crary-15}.

\subsection{Event Structures and Pomsets}
\labelsect{others-pomsets}

To express and reason about the subtle behaviours of more complex memory models such as the C memory model, there are works that use \emph{event structures} and \emph{pomsets} (partially ordered multisets), which can be viewed as mathematical structures that generalise execution graphs in \refsect{declar}.

Both event structures~\citep{winskel-87} and pomsets~\citep{pratt-86} were invented in the 1980s as mathematical models of general concurrency.
An early use of event structures for weak memory models is by \citet{cenc-07} in 2007, who described a small fragment of the Java memory model by combining event structures with operational and axiomatic techniques. Then from 2016 onwards, there have been many attempts to define memory models using related ideas~\cite{castellan-16,jeffrey-16,pichon-16,chakra-19,paviotti-20,pomsets,jeffrey-22,mois-22}.
The Weakestmo Semantics and its extension~\citep{chakra-19,mois-22} are multi-execution, i.e.,
they encode multiple execution graphs in a single event structure (cf. the Promising Semantics
\refsect{others-timestamp}).

%% file: other-semantics.tex

\OMIT{
\begin{definition}[Event Structures]
An \emph{event structure} $\langle E, \leq, \# \rangle$ consists of the following.
\begin{itemize}
  \item A set $E$ of events.
  \item A partial order $\leq$ on $E$, called the \emph{causality} relation.
        We write $a_1 \leq a_2$ to mean that $a_1$ \emph{causes} $a_2$, or equivalently that $a_2$ \emph{depends on} $a_1$.
  \item An irreflexive, symmetric relation $\#$ on $E$, called the \emph{conflict} relation.
        We write $a \# ab$ to mean that $a$ and $b$ are completely independent.
\end{itemize}
In addition, an event structure must satisfy the two properties below.
\begin{itemize}
  \item Every event depends on only finitely many other events.
        That is, for all $a\in E$, there are only finitely many $b\in E$ such that $b \leq a$.
  \item If $a_1$ and $b$ are independent, then any event that depends on $a_1$ is also independent from $b$.
        That is, for all $a_1,a_2,b\in E$, if $a_1 \leq a_2$ and $a_1 \# b$, then $a_2 \# b$.
\end{itemize}
\end{definition}


 By Rob:
 Essentially the parallelism is represented in a global event structure (dependency graph), which includes the statically-determined dependencies within the process, and includes per-execution dependencies based on which write instructions carry the value of dependent read instructions. As with the axiomatic style, cycles in a dependency graph typically indicate an invalid execution.

\begin{example}[SB: Partial ordering on events]
Recall the Store Buffer example,
without fences, \ie 
$(x \asgn 1 \scomp r_1 \asgn y) \parallel (y \asgn 1 \scomp r_2 \asgn x)$.
Under TSO there are no program-order induced dependencies within each process, however, if $r_2$ reads the value 1 for $x$ then there must appear an edge
from $x \asgn 1$ to $r_2 \asgn x$ ($\ReadEvent{2}{x}{1}$) in the global partial order for that execution.
Reads with no confirming write indicate an invalid ordering, and can be discarded from analysis.
The introduction of fences induces more (static) dependencies and further constrains the structure of the graph.
\end{example}
\scottin{(0412) I believe this felt like it came a little out of nowhere when I was first reading it.}

Verification of such systems is typically based around cycle detection on reachable states, and requires a method for enumerating all possible events from the original code.

\scottin{(0412) Mentioning program loops here (as part of what these formalisms can't or struggle to represent) could be worthwhile. I believe Rob mentioned adding this in earlier as well.}
}

\subsection{Program Transformations and Instruction Reorderings}
\labelsect{others-reordering}

\newcommand{\ppseqc}[1]{\overset{#1}{;}}
\newcommand{\ppseqcT}{\ppseqc{\mathtt{x86}}}




\newcommand{\aequiv}{&\equiv&}

A number of early works~\citep{adve-ghara-96-ieee-comput,McKenney-1,McKenney-2} specified weak memory models in terms of which pairs of instructions can be reordered.
This aligns with the workings of the processor pipeline (\refdefn{order-deps}),
and is often the level at which processors are described informally.
The first formal memory-model framework based on instruction reordering was developed by \citet{arvind-06}, who included store-atomicity as an extra criterion.

Possible instruction reordering is important in the context of compiler-related aspects of weak memory models. Common program transformations (including instruction reordering) performed by optimising compilers are not always correct under weak memory models~\cite{sevcik-08,viktor-15}.
To formulate what it means for a program transformation to be ``correct'' under a memory model, one typically develops some trace-based denotational semantics, and a transformation is then correct if the denotation of the transformed program is (a subset of) the denotation of
the original program.
Works in this line of research include \citet{sevcik-08} and \citet{burk-10} for the Java memory model, and \citet{dodds-18} and \citet{dvir-24} for the C memory model and its RA fragment.

Instruction reordering was also the basis in~\cite{colvin-21}, where the memory model is encoded as a parameter to sequential composition on the level of the program text. 
Weak memory effects are explained in terms of the standard operators for (no-parameter) sequential composition and parallel composition,
via a set of transformation rules on the instructions, in the spirit of the Concurrent Kleene Algebra~\cite{cka-09,SRA16}.
For example, instantiating the \SB litmus test on an x86 (TSO) machine,
one can derive the following.
\begin{eqnarray*}
	(x \asgn 1 \ppseqcT r_1 \asgn y)
	\pl
	(y \asgn 1 \ppseqcT r_2 \asgn x)
	\aequiv
	x \asgn 1 \parallel r_1 \asgn y
	\parallel
	y \asgn 1 \parallel r_2 \asgn x
	\\
	(x \asgn 1 \ppseqcT \codefence \ppseqcT r_1 \asgn y) \parallel (y \asgn 1 \ppseqcT \codefence \ppseqcT r_2 \asgn x) 
   \aequiv
   (x \asgn 1 ; r_1 \asgn y) \parallel (y \asgn 1 ; r_2 \asgn x) 
\end{eqnarray*}
Treating the programs on the right-hand side as under \SeqCst, one can observe that on the first line, the \Loads $r_1 \asgn y$ and $r_2 \asgn x$ can execute before the \Stores.
Meanwhile, when fences are inserted as on the second line, instances of the x86-specific sequential composition are equivalent to standard sequential composition.

This reasoning framework has been used as the basis for addressing security vulnerabilities related to speculative execution \cite{colvin-19} and information flow \cite{CoughlinInfoFlow}.
By transforming a program under a weak memory model to a program under \SeqCst, standard reasoning methods and tools become amenable, saving the need for specialised logics and techniques.
However, in contrast to timestamped messages, these program-level transformations cannot explain non-multicopy-atomic behaviours \cite{lahav-vaf-16-reorder}.


\OMIT{
Hence, if a memory model is specified in terms of possible reorderings, then we can replace certain sequential compositions by parallel compositions.
Instead of thinking about the set of related programs, one only needs to deal with one modified program with nested parallelism, to be run under \SeqCst.
This is the viewpoint adopted by \citet{colvin-21}, who represented the parallelism directly in the structure of the program text, in the spirit of the Concurrent Kleene Algebra \cite{cka-09}.
This viewpoint is illustrated in Example~\ref{ex:reorder}.

\newcommand{\TSOmm}{{\textsc{tso}}}
\newcommand{\roM}[1]{\overset{#1}{\Lleftarrow}}
\newcommand{\roT}{\roM{\TSOmm}}
\newcommand{\ffence}{\mathbf{fence}}
\newcommand{\nroT}{\not{\roM{\TSOmm}}}

\begin{example}
\label{ex:reorder}
Recall the store buffer example:
\begin{center}\begin{tabular}{l||l}
   $x \asgn 1 $ & $ y \asgn 1$ \\
  $r_1 \asgn y$ & $r_2 \asgn x$
\end{tabular}\end{center}
In the first process, since $x \asgn 1$ is independent of $r_1 \asgn y$ under TSO, these instructions could be executed in either order, as if they occur in parallel.
The same applies for the instructions of the second process.
Overall, under TSO, the store buffer program is equivalent to
\scottin{(0412) I tried to adjust the wording here to avoid two uses of under next to each other but I'm not entirely sold on it. Feel free to modify!}

An interleaving semantics shows that it is straightforward to find a trace of the program where $r_1 \asgn y$ and $r_2 \asgn x$ are the first instructions
executed, resulting in a final state where $r_1 = r_2 = 0$, which is not possible under \SeqCst.

Let the $\ffence$ be an instruction that cannot be reordered with anything else.
Then after inserting a $\ffence$ between the two instruction on each side, one can directly observe that no reordering is possible, and hence only sequentially consistent behaviours are possible.
\end{example}

Another advantage of the reordering style is that standard technique for \SeqCst becomes available, once the set of possible reorderings can be efficiently handled. 
An example is \citet{coughlin-21}, who utilized the Rely-Guarantee logic (see \refsect{tools-logic}) after handling the possible reorderings.
However, the reordering style only covers the weak memory effects that are related to pipeline reorderings. If a memory model lacks multicopy atomicity, then extra layers and mechanisms are required, as in \citet{colvin-18} and \citet{coughlin-22b}.
}

%% file: verification-and-tools.tex


\OMIT{
For the hardware designer, notable tools include Herd~\citep{alglave-14}, PipeCheck~\citep{PipeCheck}, PipeProof~\citep{PipeProof}, CheckMate~\citep{CheckMate,CheckMate-2019}, and TriCheck~\citep{TriCheck}.

The first is \emph{theorem proving}, typically using a program logic, sometimes carried out in interactive proof assistants. 
The other approach is \emph{model checking}, where automatic tools are developed to check every behaviour of a program or system exhaustively.

We now survey the techniques and tools for establishing properties of programs running under weak memory models.
\refsect{tools-logic} covers program logics, and
while requiring significant expertise, the produced results are typically general. 
\refsect{tools-check} covers model checkers.
which typically require less user interaction but produce less general results.
\refsect{tools-algo} concerns related algorithmic problems such as consistency checking, and summarises results on decidability and complexity.
}

\section{Program Logics and mechanised support}\labelsect{tools-logic}

There are three prominent families of program logics for concurrent programs that have been adapted to 
reason about weak memory effects:
\emph{Owicki-Gries} (OG) \cite{OwickiGries76},
\emph{Rely-Guarantee} (RG) \cite{Jones-RG1,Jones-RG2}, and
\emph{Concurrent Separation Logic} (CSL) \cite{SepLogicOHearn,SepLogicReynolds,CSL-16}.
%
%
This section surveys these adaptations; a summary is given in \reftable{program-logics}.

\begin{table}[t]
  \centering
  \caption{Program logics for weak memory models}
  \label{table:program-logics}
  \vskip -4mm
  \begin{tabular}{|l|l|l|l|l|}\hline
    Logic & Formalism & Memory models & Case studies & References \\\hline\hline
    OG/RG & Axiomatic & C fragment & RCU & \cite{lahav-15} \\\hline
    RG & Operational & x86-TSO & Simpson's & \cite{ridge-10} \\\hline
    RG & Operational & (parameterised) & litmus tests & \cite{abe-16} \\\hline
    OG & View-based & C fragment & Peterson's, RCU & \cite{dalvandi-20-og,dalvandi-22-og,doherty-22-unify,wright-21-og} \\\hline
    RG & potential-based & Causal Consistency & litmus tests, Peterson's & \cite{lahav-23}
    \\\hline\hline
    CSL & Axiomatic & C fragment & MS queue and others & \cite{doko-16,doko-17,vafeiadis-13,turon-14} \\\hline
    CSL & Operational & C fragment & MS queue and others & \cite{kaiser-17} \\\hline
    CSL & Operational & Rust & Rust library & \cite{rust-lang} \\\hline
    CSL & Operational & OCaml & locks, bounded queue & \cite{mevel-20,mevel-21}
    \\\hline\hline
    RG & Reordering & hardware models & Chase-Lev, Peterson's & \cite{coughlin-21,coughlin-22b} \\\hline
    Invariant & Axiomatic & (parameterised) & PostgreSQL & \cite{alglave-13,alglave-cousot-17} \\\hline
    Invariant & Operational & C fragment & Peterson's & \cite{doherty-19} \\\hline
  \end{tabular}
\end{table}

\paragraph{Axiomatic OG and RG}
Based on an axiomatic formulation of the release-acquire fragment of the C memory model, a rely-guarantee logic was developed and applied to the read-copy-update synchronisation mechanism~\cite{lahav-15}.

\paragraph{Operational OG and RG}
A rely-guarantee logic on the x86 assembly instructions was developed in \cite{ridge-10}, based on the operational formulation of x86-TSO~\cite{owens-09}. 
This logic was applied to Simpson's four-slot algorithm, and mechanised in HOL.
The rely-guarantee logic of \cite{abe-16} works on general memory models that are specified operationally, with case-studies performed on litmus tests.

Another line of research~\cite{dalvandi-20-og,dalvandi-22-og} devised a view-based formalisation (see \refsect{others-timestamp}) of the release-acquire-relaxed fragment of the C memory model, and built an Owicki-Gries logic based on this formulation. 
The main case-studies, carried out in Isabelle/HOL, are Peterson's algorithm and the read-copy-update mechanism;
related works in this line of research include~\citep{doherty-22-unify,wright-21-og}.

Similar to view-based semantics is the ``potential-based'' semantics, which has been used as the basis of a rely-guarantee logic for Causal Consistency~\cite{lahav-23}.

\paragraph{Axiomatic CSLs}
One line of research consists of the following three generations of logics: RSL (Relaxed Separation Logic)~\citep{vafeiadis-13}, GPS (Ghost states, Protocols, and Separation)~\citep{turon-14}, and FSL (Fenced Separation Logic)~\citep{doko-16,doko-17}. All three are based on the axiomatic C memory model~\citep{batty-11}, and are mechanised in Rocq (formerly Coq). 
The case-studies of these logics include simple litmus tests, while GPS in particular was applied to more realistic concurrent data structures such as the Michael-Scott queue, the circular buffer, and the bounded ticket lock.

\paragraph{Operational CSLs}
The Iris framework~\citep{jung-15,jung-18} uses an underlying operational semantics to unify various concurrent separation logics from the setting of \SeqCst.
Iris is mechanised in Rocq, which is also used by its derivatives listed below.

Iris was extended to the weak memory setting by combining with GPS. The results are iGPS~\citep{kaiser-17} and iRC11~\citep{rust-lang}; see also \citep{SepLogic-for-Promising}.
The case-studies of iGPS are the same as those of GPS, while the case-studies of iRC11 are drawn from a library of the Rust programming language.

Cosmo~\citep{mevel-20} is another CSL based on Iris for reasoning about multicore OCaml and its operationally defined memory model.
The case-studies are a spin lock, a ticket lock, and Peterson’s algorithm. 
The sequel~\citep{mevel-21} then applied Cosmo to verify a concurrent bounded queue. 
See~\citep{ferreira-10} for an alternative approach to CSL.

\paragraph{RG Based on Reorderings}

There is a rely-guarantee logic~\cite{coughlin-21} for multicopy-atomic hardware memory models, using a reordering
formalism~\cite{colvin-18,colvin-21}.  The case-study is the Chase-Lev double-ended queue for Arm, with the proof mechanised in Isabelle/HOL.  The
same authors later extended their framework to handle memory models without multicopy atomicity~\citep{coughlin-22b}, and verified
Peterson's algorithm in Isabelle/HOL.  The CLH fine-grained lock, as used in the seL4 microkernel \cite{seL4-Verification}, is verified for Arm using a related
rely-guarantee-based approach in \cite{seL4-CLH-wmm-24}.

\paragraph{Others}
There are other proof methods for invariant-based reasoning~\citep{alglave-13,alglave-cousot-17,doherty-19}.
The case studies in~\citep{alglave-13,alglave-cousot-17} are drawn from PostgreSQL, whereas in~\cite{doherty-19} the case study is Peterson's mutual exclusion algorithm.


\section{Model checking and computational problems}\labelsect{tools}

This section surveys the techniques and tools for model-checking programs that run under weak memory models.
\refsect{tools-check} covers model checkers.
\refsect{tools-algo} concerns related algorithmic problems such as consistency checking, and summarises results on decidability and computational complexity.

\subsection{Model Checkers}\labelsect{tools-check}

\begin{table}[t]
  \centering
  \caption{Model checkers for weak memory models (grouped by underlying formalisms)}
  \labeltable{model-checkers}
  \vspace{-4mm}
  \begin{tabular}{|l|l|l|}\hline
    Name & Type & Memory models 
    \\\hline\hline
    CBMC extended~\cite{CBMC-partial-orders} &
    BMC &
    x86, Power 
    \\\hline
    Herd~\cite{alglave-21} & simulator & parametric (Cat language) \\\hline
    SATCheck~\cite{demsky-15} & SMC & TSO 
    \\\hline
    CDSCheck~\cite{norris-13,norris-16} & SMC & C \\\hline
    Dartagnan~\cite{dartagnan-22} & BMC & parametric (Cat language) \\\hline
    GenMC~\cite{kokolo-21} & SMC & parametric \\\hline
    Zord~\cite{fan-23} & BMC & TSO, PSO \\\hline\hline
    McSPIN~\cite{abe-17} & BMC & parametric \\\hline
    Nidhugg~\cite{abd-16,abd-17} & SMC & TSO, PSO, Power \\\hline
  \end{tabular}
\end{table}

In the weak-memory realm, most model checkers employ either Stateless Model Checking~\citep{stateless-model-checking} or Bounded Model Checking~\citep{bounded-model-checking}. 
Here, we classify the prominent model checkers into
those based on axiomatic formalisms (\refsect{declar}),
and, less commonly, those based on operational formalisms (\refsect{others-timestamp}).
\reftable{model-checkers} summarises these model checkers, together with the memory models they support.
In the text below, we also cover the case-studies to which these model checkers have been applied.

\paragraph{Based on Axiomatic Formalisms}
Model checking under weak memory models is mostly based on axiomatic formalisms.
The most prominent axiomatic formalism is the Cat language, with the associated Herd tool~\cite{alglave-14}.
Herd simulates the possible behaviours of assembly programs, under memory models specified in Cat.
Herd and Cat have been applied to thousands of litmus tests to check whether a specification of a memory model does agree with the hardware behaviour.
They were used by Arm as the basis for developing the specifications of its memory models~\citep{alglave-21}. 
Herd is included in the ``diy tool-suite''~\cite{herd7website}, which includes facilities for automatically generating litmus tests and running them directly on hardware.

The CBMC bounded model checker \cite{CBMC} was extended to handle the memory models of x86 and Power~\citep{CBMC-partial-orders}.
Apart from litmus tests, the case-studies include components of system software, such as a queue mechanism of the Apache HTTP server, the worker synchronisation mechanism in PostgreSQL, and the read-copy-update mechanism of the Linux kernel.

Dartagnan~\citep{dartagnan-18,dartagnan-19,dartagnan-22} is a bounded model checker, where users specify memory models in the Cat language.
The benchmarks are several mutual exclusion algorithms, including Dekker's algorithm, Peterson's algorithm and the \tw{qspinlock} of the Linux kernel~\citep{haas-23,paolillo-21}.

Another recent bounded model checker is Zord~\citep{fan-23}, which is based on a novel axiomatic formalism that supports TSO and PSO, and has been applied to the SV-COMP 2020 benchmarks.

Meanwhile, there are a few stateless model checkers that are based on axiomatic formalisms.
For instance, CDSCheck~\citep{norris-13,norris-16} is based on the axiomatic C memory model~\cite{batty-11}, and uncovered an error in an implementation of the Chase-Lev double-ended queue.

There is also the line of stateless model checkers HMC~\citep{kokolo-20}, GenMC~\citep{kokolo-21} and WMC~\citep{mois-22}. They have been applied to synthetic litmus tests, mutual exclusion algorithms, and concurrent data structures (Chase-Lev double-ended queue, Treiber stack, and Michael-Scott queue).


\paragraph{Based on Operational Formalisms}
There are comparatively fewer model checkers that are based on operational formalisms to encode the weak memory models.

McSPIN~\citep{abe-17} is an extension to the SPIN model checker, and users specify memory models by constraining the operational model in McSPIN.
The case-studies include litmus tests from the documentation of Unified Parallel C and Itanium, as well as the double-checked locking algorithm and Dekker's algorithm.

Nidhugg~\citep{abd-16,abd-17} uses stateless model checking based on a specialised operational formalism of memory models.
The memory models supported are TSO, PSO, Power, and a subset of Arm.
Nidhugg accepts standard C input, and has been applied to classical benchmarks, such as Dekker’s, Lamport’s and Peterson’s mutual exclusion algorithms.

Abstract interpretation is another approach to automatic verification, but has received relatively little attention. The few existing works focus on TSO and PSO~\cite{dan-13,dan-17,suzanne-16}.

\subsection{Algorithmic Problems}\labelsect{tools-algo}

At the heart of model checkers are computational problems for deciding whether a program satisfies certain properties.
The decidability and computational complexity of these problems are substantial topics in their own right \cite{Bouajjani2024}.
This section surveys the works that address these problems under weak memory models, with the results summarised in \reftable{wmm-algo}.

\begin{table}[t]
  \centering
  \caption{Computational problems and complexity results related to weak memory models
  }
  \label{table:wmm-algo}
  \vspace{-4mm}
  \begin{tabular}{|l||ll|l|ll|}\hline
    Model
    & Consistency
    & (data-ind.)
    & Robustness
    & Reach.
    & (param.) \\\hline
    SC
    & NP-com~\citep{gibbons-97}
    & FPT~\citep{chini-20}
    &
    &
    & \\\hline
    Causal
    & NP-hard~\citep{bouaj-17,furbach-15}
    & P~\citep{bouaj-17}
    &
    &
    & \\\hline
    TSO
    & NP-hard~\citep{furbach-15}
    & FPT~\citep{chini-20}
    & PSPACE-com~\citep{bouaj-11}
    & dec~\citep{atig-10}
    & PSPACE-com~\citep{abd-20} \\\hline
    PSO
    & NP-hard~\citep{furbach-15}
    & FPT~\citep{chini-20}
    &
    & dec~\citep{atig-10}
    & \\\hline
    RMO
    & NP-hard~\citep{furbach-15}
    & FPT~\citep{chini-20}
    &
    & undec~\citep{atig-10}
    & \\\hline
    Power
    &
    &
    & PSPACE-com~\citep{derev-14}
    &
    & \\\hline
    RA
    &
    & FTP~\citep{tunc-23}
    &
    & undec~\citep{abd-19b}
    & PSPACE-com~\citep{krishna-22} \\\hline
  \end{tabular}
\end{table}

\paragraph{Consistency Checking.} 
The problem of \emph{consistency checking} is to decide whether a concurrent program yields an execution that conforms to the given memory model.

This problem is mostly NP-hard.
Indeed, when phrased in terms of axiomatic formalisms, only the program order (\po) is inferred from program text, and one needs to explore the different choices for the coherence order (\co) and the reads-from (\rf) relation (see \refsect{declar}).

\citet{gibbons-97} first showed that consistency checking under \SeqCst is NP-complete. 
Later, Cantin et al.~\cite{cantin-03,cantin-05} showed that it is NP-complete to check if a concurrent program yields a coherent execution (\refdefn{coherence}).
As a corollary, consistency checking is NP-hard under most memory models, since most memory models satisfy the coherence property.
Indeed, \citet{furbach-15} established the NP-hardness of consistency checking under the memory models covered by \citet{steinke-nutt-04}.

To circumvent the NP-hardness, one can aim for less general solutions. For instance, \citet{hangal-04} developed an incomplete but polynomial-time tool for consistency checking under TSO.

Another way to deal with the NP-hardness is to add extra premises to the problem, such as the \emph{data-independence} assumption: every value that is ever written is unique.
This means that the reads-from relation (\refdefn{reads-from}) is already supplied, as every \Read ever only has one possible \Write as its source in the reads-from relation.
Hence, one only needs to assign a suitable coherence order before checking the axioms, which makes consistency checking more tractable at times.
For instance, though consistency checking under Causal Consistency is NP-hard in general, \citet{bouaj-17} showed that it is in P when data-independence is assumed.

NP-hardness can also be approached using more fined-grained notions of complexity, namely ``fixed-parameter tractability'' (FPT).
Under SC, TSO and PSO and assuming data-independence, \citet{chini-20} showed that consistency checking is FPT: when the number of \Writes is fixed, then the problem is polynomial in the number of \Reads.

One can also parameterise the problem into the number of memory events and the number of threads. Also assuming data-independence, \citet{tunc-23} found near-linear parameterised algorithms for consistency checking under several variants of the C memory model.

\paragraph{Robustness.}
Given a memory model M, a program is said to be \emph{not robust} under M if it exhibits any extra behaviour that is not possible under SC. 
The robustness problem for a memory model M takes a program $p$ as the input, and computes whether $p$ running under M yields any extra behaviour that is not possible when $p$ runs under \SeqCst.

\citet{burnim-11a,burnim-11b} implemented techniques to test robustness under TSO and PSO.
The robustness problems for TSO and Power were shown to be PSPACE-complete~\cite{bouaj-11,derev-14}.

The robustness problem compares a weak memory model with SC. More generally, one can study the automatic comparison between any two weak memory models~\cite{wickerson-17,kokolo-23}.

\paragraph{Reachability.}
Also related to model checking (\refsect{tools-check}) is the \emph{reachability problem}, which is to decide whether a given program has an execution from a given starting state to a given final state. 

Under RMO the reachability problem is undecidable, but under TSO and PSO it is decidable, albeit beyond the primitive recursive (PR) class and has a very high computational cost~\cite{atig-10}.
In light of this computational hardness, there were works that attempted to trade completeness with efficiency under TSO~\cite{bouaj-15-lazy}.

The complexity of the reachability problem was also studied under the setting of parameterised verification. Under the weak memory models TSO and RA, the parameterised reachability problems were shown to be PSPACE-complete~\cite{abd-20,krishna-22}.

%% file: future-directions.tex

\newcommand{\Meaning}[1]{\llbracket #1 \rrbracket}
\newcommand{\MeaningM}[2]{\Meaning{#2}_{#1}}
\newcommand{\Meaningm}[1]{\MeaningM{\mm}{#1}}

\newcommand{\realscomp}{\scomp\scomp}
\newcommand{\xfence}{\texttt{fence}}

\newcommand{\labelprop}[1]{\label{prop:#1}}
\newcommand{\refprop}[1]{property~(\ref{prop:#1})}
\newcommand{\refProp}[1]{Property~(\ref{prop:#1})}

\section{Future Directions: Common Properties of Weak Memory Models}
\labelsect{future-properties}

As surveyed in \refsect{formalisms}, there are many different representations that may be used to define memory models.
With some exceptions noted below, generally missing from the literature are agreed-upon results or properties that formalisms should exhibit;
in this section we outline how the field could develop in this respect.

\subsection{Empirically-Established Properties}

The influential ``herding cats'' work of \citet{alglave-14} robustly and rigorously tests weak memory models against
litmus tests.  This work included testing real hardware across a range of architectures and vendors, and identified a
bug in a processor that resulted in a ``Programmer Advice Notice'' \cite{ReadAfterReadHazardNotice2011},
and has been used in the development of Arm processors \cite{alglave-21} and the Linux kernel \cite{alglave-18}.  

Based on this work
Arm has led the way in formalising aspects of its architecture \cite{alglave-21,alglave-24, fox-23}, including publishing
the memory model \cite[Chapter B2]{ARMv8A-Manual} 
and the \texttt{herdtools} suite of applications \cite{herd7website,AlglaveBlog}.
Ideally other companies and committees would follow suit, publishing, preferably, a formal model,
or at least a representative set of litmus tests (small concurrent programs with an associated reachability condition 
\refsect{litmus-tests})
that cover both allowed (weak) and forbidden behaviours.
This would support easier comparison across models and formalisms, and aid cross-compilation.

However, when focussing purely on empirical evaluation,
it is unclear how to achieve complete coverage of all behaviours.
This is obviously a difficult problem: given say 10 individual instruction types (loads, stores, fence types, memory ordering constraints, branch points, etc.), across just 2 concurrent processes each with 5 lines of code,
there are approximately 10 billion possible litmus tests.  This number does not include any attempt to generate the expected results.
Furthermore, litmus tests typically do not cover loops, or non-trivial conditional statements.
A ``minimal'', maintained set of tests and expected results could be kept by an architecture vendor; programmers and formalists together could access this database,
with both small litmus tests and larger programs explaining weak behaviours that can arise.
This would help formalists check that their definitions and representations are correct, and, with the 
help of a common notation that abstracts from processor-specific syntax (e.g., \cite{LISA-16,alglave-cousot-17}),
would aid comparison and re-use of results across different architectures and languages.

\subsection{Formally-Derived Properties}

Even if a representative set of tests can be developed, however,
this will not address all concerns for software development.
Programmers want to know the required solution to their programming problem -- where to place fences or constraints to maintain the logical integrity of their concurrent algorithm --
without needing to search through a database of litmus tests or to understand complex memory model specifications.

One theoretical aspect upon which the language-design community has more or less reached a consensus is that ``programs that contain no data-race should behave as if under \SeqCst''~\cite{adve-boehm-10-commun-acm},
a property known as \emph{Data-Race-Free implies \SeqCst}, or the \emph{Data Race Freedom Guarantee (DRF)}.
This relatively early concept \cite{adve1992sufficient,AdveHill93}
provides a base-level property for programmers of sequential, non-interfering code, 
similar to the principals underlying the circumstances under which localised pipelined reordering can occur
(\refsect{pipelining}).
However, the result does not necessarily help programmers of low-level systems code that \emph{does} include data races, such as system-wide locks \cite{alglave-18,seL4-CLH-wmm-24}.

What would complement such properties and litmus test conformance is a set of more general properties of programs that can be understood locally \cite{bargmann-lifting-24};
for instance, if a fence is inserted between instructions $\aca$ and $\acb$, can standard ordering be assumed?
If we let `$\realscomp$' represent strict ordering, not subject to memory model effects,
then this property of fences can be expressed as 
follows.
\begin{equation}
	\labelprop{fence-sc-cmd}
	( \aca \scomp \xfence \scomp \acb ) = 
	( \aca \realscomp \acb ) 
\end{equation}
This is the sort of property software developers need to understand their code: it does not mention any particular representation,
and does not require a global analysis of the behaviour of other threads.
We may generalise such properties as follows,
where $\Meaning{c}_\mm$ represents the set of traces that program $c$ 
generates -- in any formalism of memory model $\mm$ -- and where `$\cat$' represents concatenation of traces
and $\interleave$ represents interleaving of traces.
\begin{eqnarray}
	\Meaningm{c_1 \scomp \ffence \scomp c_2} 
	&=&
	\Meaningm{c_1} \cat \Meaningm{c_2}
	\labelprop{fence-sc}
	\\
	\Meaningm{c_1 \scomp c_2} 
	&=&
	\Meaningm{c_1} \interleave \Meaningm{c_2}
	\qquad 
	\text{if $c_1$ is \emph{independent} of $c_2$}
	\labelprop{indep-pl}
	\\
	\MeaningM{\SCmm}{c} 
	&=&
	\Meaning{c}
	\labelprop{sc-sc}
\end{eqnarray}
\refProp{fence-sc} would provide the underlying justification for the earlier program-level understanding of
\refprop{fence-sc-cmd},
giving assurance that the insertion of a (full) fence allows reasoning about the separated parts of a program separately (\ie compositionality).
\refProp{indep-pl}, for some model-specific definition of ``independent'', indicates that low-level parallelism is a factor despite the use of sequential composition.
\refProp{sc-sc} states that, for a given formalism, instantiating the memory model by the simplest -- \SeqCst --
reduces the program to a standard trace semantics, represented by $\Meaning{c}$.
Even this relatively simple property -- a sanity check on the definitions and structures used in a formal semantics -- is rarely formally proved.
A general class of properties of programs at this level would make the task of comparing formalisms more straightforward,
and provide a relatively stable basis for programmers and verifiers to understand their task.
Reduction to standard operators also opens up the possibility for existing tools and techniques to be applied for reasoning,
reducing the need for specialised inference systems~\citep{alglave-cousot-17,lahav-15,coughlin-21,doherty-19,doherty-22-unify,wright-21-og},
and expanding the range of properties and domains that can be addressed.

\section{Current and Future Directions: adapting and contributing to software development}\labelsect{concl-others}
\labelsect{future-adapt}

We outline some other research topics related to weak memory models, where
software developers must face the consequence of weak memory effects.
This includes ongoing work into verification methods and tool support, along with model checkers
(\refsects{tools-logic}{tools}); in addition, formalisation of weak memory models can and should also impact 
other aspects of formal concurrency research, such as 
verified compiler efforts (e.g. \cite{CompCertTSO-13,TriCheck}) and data race detection \cite{adve1991detecting,kasikci-15}.
These topics typically require some abstraction from the specific representation of weak memory
to facilitate more standard analyses; we give an overview of some of these below where not previously discussed.


\subsection{Properties of Programs}

\paragraph{Linearisability}  Linearisability is a widely-accepted correctness criterion for concurrent data structures. 
In contrast to \emph{reachability} of final states, linearisability describes the continual interaction of processes operating on a shared data structure.
Based on its original definition (which assumed \SeqCst), Travkin \etal~\cite{travkin-13-lineariz,travkin-14-lineariz} developed techniques to reduce a TSO program into an SC program, enabling one to utilise existing tools such as the SPIN model checker to test linearisability.
Other approaches to adapting linearisability for application to weak memory models propose alternative versions of linearisability~\citep{derrick-14,derrick-17,object-refine-1,object-refine-2}.

\paragraph{Progress/Liveness}
Most existing formalisms and tools for weak memory models handle safety properties only. Verifying \emph{liveness} properties (also called \emph{progress} properties) is more challenging even in the \SeqCst setting, with one reason being the need to address various \emph{fairness} assumptions. In the weak-memory setting, works that deal with liveness properties only started to emerge recently \cite{lahav-21,abd-23,abdulla-2024}.

\paragraph{Security Vulnerabilities}  These may emerge from microarchitectural features, including through the exploitation of speculative execution and other mechanisms that cause out-of-order execution. 
The intersection of vulnerabilities and weak memory models is thought to be a potentially important area to
investigate \cite{SoK-Barthe-22}, and 
has been addressed by several authors utilising a range of different representations
\cite{CheckMate,cats-v-spectre,Disselkoen-2019,secure-specexec-isabelle-21,colvin-19,colvin-23,pipelines-cav-25}. 
The addition of security-focussed instructions, such as Capability Hardware Enhanced RISC Instructions (CHERI), also impacts
vulnerability research \cite{zaliva-cheri-24}, and will require richer pointer-level models to address in the presence of weak memory.

\subsection{Adapting Existing Code}

\paragraph{Development of Libraries and Portability.}
The specification of concurrent libraries with weak memory is a crucial topic, as concurrent libraries are the building-blocks of real-world concurrent programs. Based on formalisms for defining memory models, much recent effort has extended and applied them to specifying and verifying concurrent libraries. These works focus on various memory models, from TSO~\cite{burck-12} and Java~\cite{emmi-19} to fragments of the C memory model~\cite{batty-13,dalvandi-21,dang-22,raad-19,singh-lahav-23}.
Large-scale migration of code between hardware platforms has been recently addressed by \citet{beck-23}, who developed a tool for automatically porting x86 code to platforms with weaker memory models, such as Arm and RISC-V. Such tools utilises fence insertion and model checking techniques, and also employs heuristics to improve scalability. 

\paragraph{Fence Insertion.}
A fence insertion algorithm aims at automatically finding a minimum set of positions in an original program to insert fences, 
so that the resultant program does not have any extra behaviour when executed under a weak memory model. 
This is important in the context of compiler-design: given a program written under the assumption of \SeqCst, the compiler needs to ensure that this program still has the original intended behaviour after it is translated to the hardware-language and run under some weak memory model. This issue was first addressed by \citet{shasha-snir-88}.
\citet{lee-00} developed a fence insertion algorithm based on analysing dominators in dependence-graphs under three weak memory models: Weak Consistency, Processor Consistency and Release Consistency.
\citet{fang-03} empirically applied the same method to the memory models of actual hardware platforms, namely IBM's Power and Intel's Pentium, which had different types of fences.


For TSO and PSO, there are fence insertion algorithms based on modelling the write-buffers~\citep{abd-12-counter-example,linden-10,linden-11,linden-13} and on predicate abstraction techniques~\citep{abd-12,abd-15b,abd-15c}.
In the case of structured programs, \citet{taheri-19} developed a polynomial-time greedy algorithm for fence insertion, and also demonstrated the NP-hardness of the minimum fence insertion problem with multiple types of fences.
Many other approaches to fence insertion have been taken~\citep{alglave-14b,bender-15,burk-07,kupers-12,joshi-15,meshman-14,meshman-15}, 
with the most recent being \cite{ober-21,singh-22}. 
Finding \emph{the} minimal set remains an open problem, however, as does finding the most computationally efficient set.

\subsection{Programming Styles and Hardware Variants}

\paragraph{Transactional Memory} Transactional memory~\citep{book-transactional} defines synchronisation mechanisms that allow the programmer to specify atomic ``blocks'' of instructions. While these blocks should appear atomic, on the hardware level they can still occur in parallel. When two atomic blocks attempt to change the same variable and a clash arises, the transactional memory system can safely revert to a consistent state. 
Naturally, transactional memory has interplay with memory consistency models~\cite{chong-18,dongol-19}.
Most recently, \citet{dalvandi-22} devised a semantics of transactional memory under the release-acquire fragment of the C memory model, and developed a program logic based on their semantics.

\paragraph{Persistent Memory} Persistent memory~\citep{book-persistency} is a relatively new type of physical memory that maintains its
state even after system failures, with the aim of avoiding the loss of data. Updates to memory are held in the volatile memory first, and they are
flushed to the persistent memory in some order, during which the persistent memory may temporarily be in some inconsistent states. In the
case of a power outage, one needs to ensure that the entire memory system can be reverted back to a consistent state. These issues, related
to memory consistency and cache coherence, have been addressed by formalisations of weak-memory
models~\cite{raad-19b,raad-20,cho-21,khyzha-21,vafbila-22,klimis-24,x86-persistency-Raad-24}.



\OMIT{
We summarise and compare the capabilities of different styles of formalisms in Table~\ref{table:concl}. Each row corresponds to each style of formalisms, while the columns represent the following:
\begin{itemize}
\item
\emph{All MCA:} the formalism applies to any (reasonable) multicopy-atomic memory model, or to specific multicopy-atomic memory models.;
\item
\emph{NMCA:} the formalism applies to non-multicopy-atomic memory models;
\item
\emph{Model Check:} has publicly available model checking support;
\item
\emph{Theorem Prover:} properties checked in a theorem prover;
\item
\emph{Inference System:} the formalism support an inference system for establishing correctness.
\end{itemize}

\newcommand{\Tick}{\checkmark}
\newcommand{\Cross}{$\times$}

\begin{table}[t]
  \caption{Comparison of weak-memory formalisms}
  \label{table:concl}
  \centering
  \begin{tabular}{|l|c|c|c|c|c|} \hline
    & All MCA & NMCA & Model Check & Theorem Prover & Inference System
    \\\hline
    Herding cats \cite{alglave-21} &
    \Tick &
    \Tick &
    \Tick &
    \Cross &
    \Cross
    \\
    x86-TSO \cite{sewell-10} &
    \Cross &
    \Cross &
    \Cross &
    \Cross &
    \Cross \\
    Promising \cite{kang-17} &
    \Tick &
    \Tick &
    \Cross &
    \Tick &
    \Tick \\
    Colvin \cite{colvin-21} &
    \Tick &
    \Cross &
    \Tick &
    \Tick &
    \Tick \\\hline
  \end{tabular}
\end{table}
}

%% file: conclusions.tex

\section{Conclusions}\labelsect{concl}

Weak memory models are in a sense historical legacies from optimisations made by hardware and compilers 
for single-core architectures (\refsect{history}).  Computational efficiences can be gained in some cases by reordering 
the execution of individual statements of code, with a minimum requirement being simply that the programmer
will never know that the reordering has taken place.  However, these changes, already somewhat exposed
to programmers of concurrent code with agressive compiler optimisations, became significant with the
advent of multicore architectures, initiating the first wave of attempts at formalising these effects (\refsect{prelim}).

Often this issue can be addressed by having programming abstractions between the assembler
and the high-level code; however this is not possible for programmers at the interface
of hardware and software, for instance, operating systems developers \cite{alglave-18}.
In this context safety, security and performance are all of high importance, and a clear place where
a rigorous formal model of the behaviour of multicore systems is invaluable \cite{SoK-Barthe-22}.
Since low-level details are almost unavoidable when addressing weak memory model concerns, further refinement of
formalisms to include detailed instruction semantics \cite{Sail-19} would assist binary-level analysis (e.g.,
\cite{lasagne-22,risotto-22}).

Approaches to formalising weak memory effects range from modelling features of the architectures directly 
(e.g., \refsect{operat}) through to abstracting away from such details to axioms of global traces (the axiomatic approach, \refsect{declar}).
Within these broad classifications many other representations are used (\refsect{formalisations}),
commonly including a \emph{partial order} on events:
a convenient representation for defining order between events where necessary,
while leaving order unspecified when weak memory effects manifest.
Many are supported by verification techniques (\refsect{tools-logic}) and model checking tools (\refsect{tools});
however what is lacking from the field are rigorous sets of common properties, whether they be determined
by small litmus tests or program-level statements of behaviour (\refsect{future-properties}).


Researchers at Arm have adopted the axiomatic style to specify their model,
which is made concrete
via conformance to a large set of litmus tests.  
This essentially makes the axiomatic style the \emph{de facto} standard in describing weak memory models.
Typically new semantics styles are validated by showing conformance to the expected outcomes of a set of litmus tests,
of which canonical sets exist for both Arm and RISC-V.
The weakness of the axiomatic style is that it does not lend itself directly to a straightforward semantics for \emph{programs}.
While memory models can be compared, and general properties of memory models described~\cite{alglave-14},
one cannot easily take an abstract program and apply a set of (denotational) rules to obtain the set of traces for a program
written in some abstract syntax.
This is addressed by, for instance, the Promising semantics~\citep{kang-17},
an operational semantics which includes axiomatic-style relations.
However this semantics, partly due to capturing a wide-range of weak behaviours,
is quite complex to apply to even simple programs.
Explanations based on microarchitectural features are usually more straightforward to describe and implement, but less general and less transferrable 
to the software level.

This leaves quite a gap between hardware developers, compiler writers, and eventually, software programmers.
Low level systems code needs clear guidelines to get the most out of modern hardware capabilities, since
the overuse of fences may lead to performance downgrades.
There are also many other abstractions which would benefit from representations
that are amenable to further analysis (\refsect{future-adapt}), not tied too directly to weak memory
notions.
The field would benefit greatly from language and hardware designers talking more in terms of desirable
properties, rather than representations.

\OMIT{

\begin{table}[h]
  \centering
  \caption{Main formalisms for the main memory models}
  \begin{tabular}{|c||c|c|c|} \hline
  platform & decl. & mech. & other \\\hline\hline
  x86      & \cite{owens-09} 
           & \cite{owens-09} 
           & \cite{colvin-21} \\\hline
  Power    & \cite{IBM-POWER-93,adir-03}
           & \cite{sarkar-11}
           & \\\hline
  Arm      & \cite{alglave-21,pulte-17}
           & \cite{pulte-17,ModellingARMv8}
           & \cite{colvin-21} \\\hline
  RISC-V   & \cite{RISC-V}
           &
           & \\\hline\hline
  Java     & \cite{manson-05}
           & \cite{jagad-10}
           & \\\hline
  C        & \cite{batty-11}
           & \cite{nienhuis-16}
           & \cite{colvin-23} \\\hline
  \end{tabular}
\end{table}
}


\OMIT{
We summarize and compare the capabilities of different styles of formalisms in Table~\ref{table:concl}. Each row corresponds to each style of formalisms, while the columns represent the following:
\begin{itemize}
\item
\emph{All MCA:} the formalism applies to any (reasonable) multicopy-atomic memory model, or to specific multicopy-atomic memory models.;
\item
\emph{NMCA:} the formalism applies to non-multicopy-atomic memory models;
\item
\emph{Model Check:} has publicly available model checking support;
\item
\emph{Theorem Prover:} properties checked in a theorem prover;
\item
\emph{Inference System:} the formalism support an inference system for establishing correctness.
\end{itemize}

\newcommand{\Tick}{\checkmark}
\newcommand{\Cross}{$\times$}

\begin{table}[t]
  \caption{Comparison of weak-memory formalisms}
  \label{table:concl}
  \centering
  \begin{tabular}{|l|c|c|c|c|c|} \hline
    & All MCA & NMCA & Model Check & Theorem Prover & Inference System
    \\\hline
    Herding cats \cite{alglave-21} &
    \Tick &
    \Tick &
    \Tick &
    \Cross &
    \Cross
    \\
    x86-TSO \cite{sewell-10} &
    \Cross &
    \Cross &
    \Cross &
    \Cross &
    \Cross \\
    Promising \cite{kang-17} &
    \Tick &
    \Tick &
    \Cross &
    \Tick &
    \Tick \\
    Colvin \cite{colvin-21} &
    \Tick &
    \Cross &
    \Tick &
    \Tick &
    \Tick \\\hline
  \end{tabular}
\end{table}
}